\documentclass{jfm}

\usepackage{graphicx}
\usepackage{newtxtext}
\usepackage{newtxmath}
\usepackage{natbib}
\usepackage{hyperref}
\usepackage{multirow}
\hypersetup{
    colorlinks = true,
    urlcolor   = blue,
    citecolor  = black,
}

\newcommand{\RomanNumeralCaps}[1]
\usepackage{soul}

\title{One-dimensional and weakly two-dimensional swash on a plane beach: Application of shock solution}

\title{Swash flow due to normal and obliquely incident bores}

\title{Swash flow due to obliquely incident bores}

\author{Hyungyu Sung\aff{1},
  Pedro Lomonaco\aff{2}, 
  Patricia Chardon-Maldonado\aff{3},
  Ryan P. Mulligan\aff{4},
  Jason Olsthoorn\aff{4},
  Jack A. Puleo\aff{5}
 \and Nimish Pujara\aff{6}\corresp{\email{npujara@wisc.edu}}}

\affiliation{\aff{1} Department of Mechanical Engineering, University of Wisconsin-Madison, Madison, WI 53706, USA
\aff{2} O.H. Hinsdale Wave Research Laboratory, Oregon State University, Corvallis, OR 97331, USA
\aff{3} Caribbean Coastal Ocean Observing System Inc., Mayaguez, PR 00680, Puerto Rico
\aff{4} Department of Civil Engineering, Queen’s University, Kingston, ON K7L 3N6, Canada
\aff{5} Center for Applied Coastal Research, University of Delaware, Newark, Delaware 19716, USA
\aff{6} Department of Civil and Environmental Engineering, University of Wisconsin-Madison, Madison, WI 53706, USA}

\begin{document}
\graphicspath{{JFM/Figures/}}
\maketitle

\begin{abstract}
We present a new solution to the nonlinear shallow water equations and show that it accurately predicts the swash flow due to obliquely approaching bores in large-scale wave basin experiments. The solution is based on an application of Snell's law of refraction in settings where the bore approach angle $\theta$ is small. We use the weakly two-dimensional nonlinear shallow water equations [Ryrie (\textit{J. Fluid Mech.}, vol. 129, 1983, p. 193)], where the cross-shore dynamics are independent of, and act as a forcing to, the alongshore dynamics. Using a known solution to the cross-shore dynamics [Antuono (\textit{J. Fluid Mech.}, vol. 658, 2010, p. 166)], we solve for the alongshore flow using the method of characteristics and show that it differs from previous solutions. Since the cross-shore solution assumes a constant forward-moving characteristic variable, $\alpha$, we term our solution the `small-$\theta$, constant-$\alpha$' solution. We test our solution in large-scale experiments with data from fifteen wave cases, including normally incident waves and obliquely incident waves generated using the wall reflection method. We measure water depths and fluid velocities using in situ sensors within the surf and swash zones and track shoreline motion using quantitative imaging. The data show that the basic assumptions of the theory (Snell's law of refraction and constant-$\alpha$) are satisfied and that our solution accurately predicts the swash flow. In particular, the data agrees well with our expression for the time-averaged alongshore velocity, which is expected to improve predictions of alongshore transport at coastlines.
\end{abstract}


%{\bf MSC Codes }  {\it(Optional)} Please enter your MSC Codes here

    \section{Introduction}
    \label{sec:Introduction}

The swash zone is defined as the region where the waterline moves up and down on the beach due to wave action, where the flow is unsteady \citep{elfrink2002hydrodynamics, chardon2016advances}, turbulent \citep{petti2001turbulence, longo2002turbulence}, and multiphase due to presence of sediment and bubbles \citep{ battjes1974surf, peregrine1983breaking, bakhtyar2009modeling}. On sandy beaches  exposed to surface waves, the swash zone is the boundary between the land and water, interacting with surf zone flows and governing beach morphodynamics. For example, the final dissipation of short frequency waves occurs in the swash zone, which serves as a driver of sediment transport \citep{brocchini2008recent, vousdoukas2014role}. The resulting beach morphology, in turn, influences both surf and swash flows, behaving as a connected system \citep{masselink2006swash}. Additionally, swash zone dynamics determine the fate of pollutants \citep{elko2022human}, particles such as plastic debris \citep{davidson2023beaching, poulain2024laboratory}, driftwood \citep{murphy2024application}, and exchanges between ocean and groundwater \citep{horn2006measurements, benoit2025groundwater}. 

The alongshore flow plays a significant role in swash zone dynamics, as a large portion of transport in nearshore areas is due to alongshore drift \citep{dean2004coastal,masselink2006swash, de2016initial,murphy2025development}. The interaction between cross-shore and alongshore flows can also enhance sediment transport \citep{austin2011alongshore}. To accurately estimate the total transport inside the surf and swash zones, proper boundary conditions related to the moving shoreline must be established. Calculating mass and momentum fluxes within the swash zone from these boundary conditions is essential, as most nearshore flow field studies utilize wave-averaged models \citep{longuet1970longshore1, longuet1970longshore2, brocchini1996integral, winckler2013advective}. Therefore, improved fundamental understanding of two-dimensional swash flows is necessary to advance modeling in nearshore hydrodynamics and transport processes.

Classical models of swash zone dynamics are derived based on the nonlinear shallow water equations \citep{peregrine1972equations}. The hyperbolic nature of these equations allows them to be expressed in terms of characteristic variables (also known as Riemann invariants). For example, \citet{carrier1958water} obtained an exact solution for the nonlinear shallow water equations for the case of a swash induced by non-breaking waves reflected from the beach based on a hodograph transformation of the characteristic variables. \citet{synolakis1987runup} extended this idea to derive the maximum run-up and breaking criteria for the climb of solitary waves on a slope. However, these approaches are not applicable when waves break. Instead, the propagation of a breaking wave, referred to as a bore, is modeled as a moving discontinuity or `shock' \citep{whitham1958propagation}. For the climb of a bore on a sloped beach, \citet{ho1962climb} and \citet{shen1963climb} derived the well known solution for one-dimensional swash near the shoreline by analyzing the asymptotic behavior of the characteristics during bore collapse. \citet{peregrine2001swash} extended this analysis to the entire swash zone, drawing on the similarity between bore-drive swash and dam-break flows. Later, \citet{ryrie1983longshore} further extended these asymptotic solutions to a weakly two-dimensional swash induced by oblique bores under the assumption of a small incident angle. Building on these theoretical foundations, various researchers have investigated run-up \citep{pedersen1983run, peregrine2001swash, pujara2015experimental}, boundary layer effects and friction \citep{pedersen1983run, hogg2004effects, chanson2009application, pedersen2013runup, pujara2015swash, pujara2016integral}, and the interaction between the surf and swash zones \citep{brocchini1996integral, brocchini2008recent}. 

Separate from these analytic solutions, \citet{antuono2010shock} introduced a quasi-analytic, one-dimensional solution for a propagating bore on a sloping beach by assigning offshore boundary conditions that maintain a constant value of the forward-moving characteristic variable of the nonlinear shallow water equations. With this boundary condition, it was possible to calculate the full flow field throughout the domain and overcome the asymptotic limitations of previous analytic solutions at the cost of loss of generality. However, it has remained unclear whether the boundary condition corresponds to a physical situation or whether they merely provide a way to obtain a closed form solution. Furthermore, for the case of obliquely incident waves, although there exist some field \citep{austin2011alongshore, puleo2020role} and laboratory \citep{schuellersurf} studies, the ability of previous solutions \citep[\textit{e.g.},][]{ryrie1983longshore} to provide realistic predictions has not been directly tested.

Here, we examine the swash flow due to obliquely incident breaking waves through a combination of theory and laboratory experiments. For the theory, we combine the approaches in \citet{ryrie1983longshore} and \citet{antuono2010shock} to obtain a new solution to the weakly two-dimensional nonlinear shallow water equations (\S \ref{sec:Theory}), using the characteristic variables to build a solution based on offshore boundary conditions and the conservations of mass and momentum across the bore. We then describe laboratory experiments in a large-scale wave basin where normal and oblique waves travel toward a fixed, impermeable sloping surface that models a beach (\S \ref{sec:Laboratory Experiments}). The comparison between our theoretical solution and laboratory data (\S \ref{sec:Comparison with theoretical solutions}) shows good agreement, where the only two free parameters in the theory are obtained from the experiments. We end with conclusions and directions for future research (\S \ref{sec:Conclusions}).

    \section{Theory}
    \label{sec:Theory}

    \subsection {Nonlinear shallow water equations}
    \label{sec:Nonlinear shallow water equations}

\begin{figure}
      \centerline{\includegraphics[width=1\linewidth]{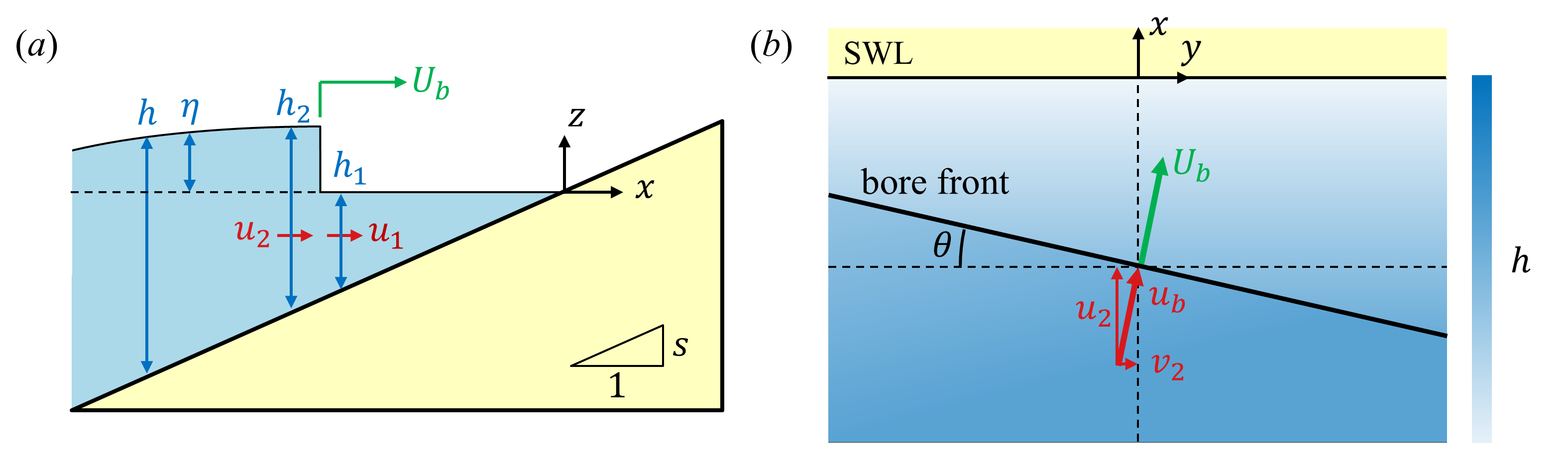}}
      \caption{Definition sketch for an obliquely approaching bore: (\textit{a}) side view; (\textit{b}) top view.}
    \label{fig:Definition_sketch_bore_diagram}
    \end{figure}

As waves approach the shoreline and travel into shallower water, the wave front steepens and the water surface behind the front flattens as the wave amplitude grows. Consequently, the length scales associated with horizontal variations of the water surface and velocity become much greater than the water depth, leading to decreased importance of frequency dispersion. Meanwhile, the growth of wave amplitude increases the importance of wave nonlinearity. Therefore, the conservation of mass and momentum are governed by the nonlinear shallow water equations (NSWEs). The two-dimensional NSWEs are
\begin{subequations}\label{eq:NSWEs}
\begin{align}
  \frac{\partial h}{\partial t} + \frac{\partial}{\partial x}(hu) + \frac{\partial}{\partial y}(hv) &=0, \label{eq:NSWEsa} \\
  \frac{\partial u}{\partial t} + u \frac{\partial u}{\partial x}+ v \frac{\partial u}{\partial y}+ \frac{\partial \eta}{\partial x} &= 0, \label{eq:NSWEsb} \\
\frac{\partial v}{\partial t} + u \frac{\partial v}{\partial x}+ v \frac{\partial v}{\partial y}+ \frac{\partial \eta}{\partial y} &= 0,  \label{eq:NSWEsc} 
\end{align}
\label{eq:NSWEsabc}
\end{subequations}
where $(x,y)$ are cross-shore and alongshore coordinates, respectively, $t$ is time, $h$ is total water depth, $\eta$ is free-surface displacement measured from still water level, and $(u,v)$ are depth-averaged velocities in the $(x,y)$ directions (see definition sketch in figure \ref{fig:Definition_sketch_bore_diagram}). The origin of the coordinate system is located at the still water line (SWL) and $h_1=h-\eta$ is the undisturbed water depth without the bore. For a constant-sloped, non-deformable beach, $h_1 = -x$ and the offshore boundary of the domain is taken to be $x=-1$. All the quantities have been made dimensionless \citep{carrier1958water} as 
\begin{equation}
    t=\frac{t^*}{t_0^*},\: (x,y)=\frac{(x^*, y^*)}{l_0^*},\: (u, v)=\frac{(u^*, v^*)}{u_0^*},\: h=\frac{h^*}{h_0^*}, \,\, \text{where} \,\, l_0^* = \frac{h_0^*}{s}, \: u_0^*=\sqrt{g^*h_0^*}, \: t_0^*=\sqrt{\frac{l_o^*}{g^*s}}, \label{eq:nondimensionalization}
\end{equation}
where the asterisk ($^*$) indicates a dimensional variable, $s$ is the beach slope, $h_0^*$ is the undisturbed water depth at the offshore boundary of the domain, and $g^*$ is the gravitational acceleration. 
    
To solve for the flow due to an obliquely approaching bore, we first consider the magnitude of the approach angle, $\theta$ (see figure \ref{fig:Definition_sketch_bore_diagram}), and make the assumption that it is small. This is not a particularly restrictive assumption since even a 30$^\circ$ approach angle, which would be considered highly oblique, has a value of $\theta \approx 0.5$ rad for which $ \theta \approx \sin \theta \approx \tan \theta$ and $\cos \theta \approx 1$ are reasonable approximations. Using the assumption of small $\theta$, we aim to simplify the NSWEs (Eqs.~\ref{eq:NSWEsabc}) to derive a weakly two-dimensional set of governing equations. We do this by first considering Snell's law, which when applied to the surf and swash zones, implies that the parameter 
\begin{equation}
    \varepsilon = \frac{\sin{\theta_0}}{U_{b0}} \label{eq:deltadef}
\end{equation}  
remains constant. Here, $\theta_0$ is the approach angle and $U_{b0}$ is the dimensionless bore speed, with the subscript $0$ denoting that both are specified at the offshore boundary of the domain. We refer to $\varepsilon$ as the effective bore approach angle since it quantifies the bore obliqueness and it is a key dimensionless parameter in the theory that follows. Since $\theta_0$ is assumed to be small and $U_{b0}$ is $O(1)$ to leading order, $\varepsilon$ is small.

Following \citet{ryrie1983longshore}, we define a pseudo-time, $\tau$, as
\begin{equation}
    \tau(t,y) = t - \varepsilon y , \label{eq:taudef}
\end{equation}  
which captures how the flow solution translates along the $y$ direction. In other words, the wave passes the same cross-shore position, $x$, at the same pseudo-time, $\tau$, regardless of the alongshore position, $y$. This effectively makes the problem one-dimensional, since the full two-dimensional solution in $(x,y,t)$ can be obtained in the one-dimensional coordinates $(x,\tau)$. 

Inserting the psuedo-time definition (Eq.~\eqref{eq:taudef}) into the governing equations (Eqs.~\eqref{eq:NSWEsabc}) and using the assumption of a small effective approach angle (\textit{i.e.}, small $\varepsilon$), we obtain 
\begin{subequations}\label{eq:NSWEs2}
\begin{align}
  \frac{\partial h}{\partial \tau} + \frac{\partial}{\partial x}(hu) &=0, \label{eq:NSWEs2a} \\
  \frac{\partial u}{\partial \tau} + u \frac{\partial u}{\partial x}+ \frac{\partial h}{\partial x} &= -1,  \label{eq:NSWEs2b} \\
\frac{\partial v}{\partial \tau} + u \frac{\partial v}{\partial x} -\varepsilon \frac{\partial h}{\partial \tau} &= 0 ,  \label{eq:NSWEs2c}  
\end{align}
\label{eq:NSWEs2abc}
\end{subequations}
which we call the weakly two-dimensional NSWEs. Here, we have chosen the cross-shore velocity to be $u = O(1)$ and the alongshore velocity to be $v = O(\varepsilon)$ in accordance with the small-$\theta$ assumption. Then, Eqs.~(\ref{eq:NSWEs2a}-\ref{eq:NSWEs2b}) are $O(1)$ and we have dropped terms of $O(\varepsilon^2)$, and Eq.~\eqref{eq:NSWEs2c} is $O(\varepsilon)$ and we have dropped terms of  $O(\varepsilon^3)$. Physically, the assumption of small effective angle implies that spatial gradients in the alongshore direction are weaker than those in the cross-shore direction. As first noted by \citet{ryrie1983longshore}, the advantage of the weakly two-dimensional NSWEs is that there is a one-way coupling from cross-shore dynamics (Eqs.~(\ref{eq:NSWEs2a}-\ref{eq:NSWEs2b})) to the alongshore dynamics (Eq.~\eqref{eq:NSWEs2c}) allowing solutions to the cross-shore problem to be used to find solutions to the alongshore flow. The weakly two-dimensional NSWEs (Eqs.~\eqref{eq:NSWEs2abc}) can be also written in characteristic form as
\begin{subequations}\label{eq:characteristics NSWEs}
\begin{align}
  \frac{d\alpha}{d\tau} &=0 \; \text{ along the curve } \ \frac{dx}{d\tau}=u+c, \quad \text{where } \alpha=u+2c+\tau,\label{eq:alpha} \\
  \frac{d\beta}{d\tau} &=0 \; \text{ along the curve } \ \frac{dx}{d\tau}=u-c, \quad\text{where } \beta=u-2c+\tau,\label{eq:beta} \\
  \frac{d\gamma}{d\tau} &=0 \; \text{ along the curve } \ \frac{dx}{d\tau}=u, \quad \text{where } \gamma=\frac{v}{\varepsilon}-h-x-\frac{1}{2}u^2,\label{eq:gamma}  
\end{align}
\end{subequations}
where $c=\sqrt{h}$ is the local wave speed. Note that these characteristics travel only along the $x$-axis due to the weakly two-dimensional nature of the equations.

    \subsection{Review of one-dimensional flow solutions}
    \label{sec:Review of one-dimensional flow solutions}

In anticipation of the two-dimensional solution in \S \ref{sec:Weakly two-dimensional flow solution}, we first review the cross-shore solutions of a normally incident bore (figure \ref{fig:Definition_sketch_bore_diagram}a) due to \citet{ho1962climb, shen1963climb, peregrine2001swash,antuono2010shock}.

The bore travels with a speed $U_b$ and the water velocity behind and in front of the bore are $u_2$ and $u_1$, respectively. The mass and momentum conservation across the bore follows Rankine-Hugoniot conditions \citep{stoker1957water}
\begin{subequations}\label{eq:bore discontinuity}
\begin{align}
  U_b[h] &=[uh], \label{eq:bore discontinuitya} \\
  U_b[uh] &= \left[\frac{h^2}{2}+u^2h\right],\label{eq:bore discontinuityb}
\end{align}
\end{subequations}
where $[f]=f_2-f_1$ indicates the jump of the quantity $f$ across the bore. 

Following \citet{antuono2010shock}, we re-write the jump conditions as
\begin{subequations}
\begin{align}
    U_b &= u_1 + \sqrt{\frac{1}{2}\left(\frac{h_2^2}{h_1}+h_2\right)},
    \label{eq:Antuono_U_b} \\
    (u_2 - u_1) &= (h_2-h_1)\sqrt{\frac{1}{2}\left(\frac{1}{h_2}+\frac{1}{h_1}\right)}.
    \label{eq:Antuono_u1 u2 relation}
\end{align}
\end{subequations}
By writing $u_1$ and $u_2$ in Eq.~\eqref{eq:Antuono_u1 u2 relation} in terms of their respective forward characteristic variables, $\alpha_i=u_i+2c_i+\tau$ (where $i=1,2$ denote the flow data either side of the bore), we can write the jump in depth across the bore as the polynomial equation
\begin{equation}
    z^6-9c_1z^4+8\sqrt{c_1}(\alpha_2-\alpha_1+2c_1)z^3-\left[ 2(\alpha_2-\alpha_1+2c_1)^2+c_1^2\right]z^2+c_1^3 = 0,
    \label{eq:Antuono_Polynomial}
\end{equation}
where $z=c_2/\sqrt{c_1}$. Since the bore is assumed to propagate into quiescent water on a beach of constant slope, we have $u_1=0$ and $h_1=-x_b$, where $x_b < 0$ is the bore position, and hence $\alpha_1$ is known. To close the problem, \citet{antuono2010shock} assumed $\alpha_2$ to be a known constant. With that, we have a closed system with three unknowns $h_2, u_2, U_b$ and three equations. We first solve Eq.~\eqref{eq:Antuono_Polynomial} for $h_2$, then obtain $u_2$ from the definition of the forward moving characteristic variable (Eq.~\eqref{eq:alpha}), and find the bore speed $U_b$ via Eq.~\eqref{eq:Antuono_U_b}. Then, the bore position $x_b$ can be obtained by numerically integrating the bore speed
\begin{equation}
    \frac{dx_b}{d\tau} = U_b(h_1,u_1,\tau), \quad x_b(0) = -1,
    \label{eq:Antuono_x_b}
\end{equation}
where $\tau=t$ for the one-dimensional (cross-shore) problem. 

The solution shows the well known phenomenon of bore collapse: As the bore approaches the shoreline ($h_1 \xrightarrow{} 0$), the water depth jump across the bore vanishes and the flow velocities $u_2$ and $U_b$ approach the same limit $U_s$, which can then be interpreted as the initial shoreline velocity of the swash flow \citep{whitham1958propagation, keller1960motion}. This shoreline is defined as the moving point where $h=0$. After bore collapse (\textit{i.e.}, after the bore cross the still water line at $x=0$), the Rankine-Hugoniot conditions are no longer applicable and the bore transitions into a moving shoreline in the ensuing swash. The velocity of this shoreline motion is given by $u_s = \alpha_2-\tau$, which is known thanks to the constant $\alpha_2$ assumption, and it can be integrated to find the shoreline position.

So far, we have calculated the bore solution: the bore trajectory before bore collapse and the shoreline motion after bore collapse. To calculate the full flow field, we note that the values of $u_2$ and $c_2$ are known from the bore solution at any bore position $x_b$, and hence we also have the resulting $\beta_2$ that is carried by the backward moving characteristics in the offshore direction. Similarly, in the swash, we have the flow variables on the moving shoreline, $u_s = \alpha_2-\tau$ and $h_s = 0$, and hence the resulting $\beta_s$. On backward moving characteristics that emanate from the bore (or the shoreline) at a point $(x_b,\tau_b)$ (or $(x_s,\tau_s)$) with characteristic variable valued $\beta_2$ (or $\beta_s$), the following equation holds at a general point $(x, \tau)$:
\begin{equation}
    x = x_b + \left(\frac{\alpha_2+3\beta_2}{4}\right)(\tau-\tau_b)-\frac{\tau^2}{2}+\frac{\tau_b^2}{2},
    \label{eq:Antuono_beta x}
\end{equation}
which allows for calculation of the value of $\beta(x,\tau)$ throughout the domain. With both $\alpha$ and $\beta$ values known, we can calculate the full flow field as
\begin{equation}
    u(x,\tau) = \frac{\alpha_2+\beta(x,\tau)}{2}-\tau, \quad c(x,\tau) = \frac{\alpha_2-\beta(x,\tau)}{4},
    \label{eq:Antuono_u c equation using alpha and beta}
\end{equation}
using again the constant-$\alpha$ assumption to close the system so that $\alpha = \alpha_2$ throughout the flow behind the bore (or the shoreline). 

Although the problem is now fully solved, it is useful to understand what a constant $\alpha(-1,\tau)=\alpha_2$ implies about the offshore boundary condition in physical terms. First, since the forward moving characteristics travel faster than the flow, applying constant-$\alpha$ at the offshore boundary is sufficient to guarantee the flow field maintains a constant $\alpha$ value throughout the domain behind the bore, as required. Second, from the definition of $\alpha$, we see that quiescent conditions at the offshore boundary at $t=0$ give $\alpha(-1,0)= 2$, and thus one measure of the strength of the incoming bore that enters the domain at $t=0$ is given by $(\alpha_2 - 2)$. Finally, the implications of the constant $\alpha_2$ on the flow variables at the offshore boundary can be understood by writing $u(-1,\tau)=u^I(-1,\tau)+u^R(-1,\tau)$ and $c(-1,\tau)=1+c^I(-1,\tau)+c^R(-1,\tau)$, where the superscript $I$ refers to the \textit{incoming} sub-critical flow and the superscript $R$ refers to the \textit{reflected} (or outgoing) flow. We see that these terms are given by 
\begin{subequations} \label{eq:Antuono_boundary condition}
\begin{align}
    u^I(-1,\tau)=\frac{\alpha_2-2-\tau}{2}, &\quad c^I(-1,\tau)=\frac{\alpha_2-2-\tau}{4}, \text{ and }\label{Antuono_boundary condition_I} \\
    u^R(-1,\tau)=\frac{\beta(-1,\tau)+2-\tau}{2}, &\quad c^R(-1,\tau)=-\frac{\beta(-1,\tau)+2-\tau}{4}.\label{Antuono_boundary condition_R}
\end{align}
\end{subequations}
Thus, at the offshore boundary, both the velocity and the square root of the depth decrease linearly in time for the incoming bore. As we show in \S \ref{sec:Laboratory Experiments}, this is not unlike the physical situation for bores approaching a coast.

Aside from the above constant-$\alpha$ solution due to \cite{antuono2010shock}, there are the well known results by \citet{ho1962climb} and \citet{shen1963climb}, who found that the shoreline motion in the swash is only dependent on the initial shoreline velocity and behaves like a particle moving freely under gravity. Through asymptotic analysis of the singularity of the characteristics as the bore approaches the shoreline, they found the shoreline motion to be described by
\begin{subequations} \label{eq:Shen Meyer}
\begin{align}
  x_s &= U_s\tau_s-\frac{1}{2}\tau_s^2,\label{eq:Shen Meyer_x_s} \\
  u_s &= U_s-\tau_s,\label{eq:Shen Meyer_u}
\end{align}
\end{subequations}
where $x_s$ is the shoreline position and $u_s$ is the shoreline velocity. Here, we have defined a new time coordinate for the swash zone $\tau_s=\tau-\tau_c$ for convenience, where $\tau_c$ indicates the time of bore collapse. 

The analytic solutions for the flow velocity and the free-surface displacement behind the moving shoreline are \citep{ho1962climb, shen1963climb, peregrine2001swash}
\begin{subequations}\label{eq:Peregrine Williams}
\begin{gather}
  h = \frac{1}{9} \left(U_s-\frac{1}{2}\tau_s-\frac{x}{\tau_s} \right)^2,\label{eq:Peregrine Williams_h} \\
  u = \frac{1}{3}\left ( U_s-2\tau_s+2\frac{x}{\tau_s}\right),\label{eq:Peregrine Williams_u}
\end{gather}
\end{subequations}
where $x>0$ is the cross-shore position above the still-water line. Note that these solutions are closely linked to a dam-break flow and also have a constant $\alpha$ behind the moving shoreline. Even though Eq.~\eqref{eq:Peregrine Williams} is an asymptotic solution near the shoreline ($(x_s - x) \ll 1$), it has been shown to predict laboratory data of swash flow very well for a large portion of the swash cycle \citep{pujara2015swash}. 

To see the link between the asymptotic solution due to \citet{ho1962climb, shen1963climb} and the constant-$\alpha$ solution due to \citet{antuono2010shock}, note that the initial shoreline velocity in the constant-$\alpha$ solution is $U_s=\alpha_2-\tau_c$, which is a known constant from the bore solution (specifically Eq~\eqref{eq:Antuono_Polynomial}). Therefore, the initial shoreline velocity $U_s$ is directly linked to the constant $\alpha_2$ and $u+2c+\tau_s=U_s$ remains invariant in the swash zone. Physically, since $h=0$ at the shoreline, all the information about the incoming flow before bore collapse is collected into $U_s$, which becomes the only parameter needed to describe the swash flow. Moreover, since $U_s$ is directly dependent on $\alpha_2$ in \citeauthor{antuono2010shock}'s (\citeyear{antuono2010shock}) solution, the swash flow can also be fully described by the constant $\alpha_2$ value in that scenario. We discuss the relationship between $U_s$ and $\alpha_2$ in further detail in \S \ref{sec:Numerically computed results}.

    \subsection {Weakly two-dimensional flow solution}
    \label{sec:Weakly two-dimensional flow solution}
    
To extend the cross-shore solution in \S \ref{sec:Review of one-dimensional flow solutions} to obliquely approaching bores, we consider a bore approaching the coast at an angle $\theta$ and with speed $U_b$ (figure \ref{fig:Definition_sketch_bore_diagram}b). The flow velocity magnitude behind the bore is denoted as $u_b$ with components $u_2$ in the cross-shore direction and $v_2$ in the alongshore direction. Momentum conservation in the direction tangential to the bore face shows that $u_b$ must be perpendicular to the bore. Therefore, $u_2=u_b \cos \theta$ and $v_2 = u_b \sin \theta$, and the forward moving characteristics in the $x$ direction behind the bore have 
\begin{equation}
    \alpha_2 = u_b\cos\theta + 2c_2 + \tau.
    \label{eq:new alpha 2}
\end{equation}
By writing $\tau$ in terms of $\alpha_1$, we get
\begin{equation}
    u_b = \sec \theta (\alpha_2-\alpha_1 + 2(c_1-c_2)).
    \label{eq:new antuono u1 u2 relation}
\end{equation}

Following a similar procedure to \S \ref{sec:Review of one-dimensional flow solutions} , we can obtain a polynomial equation for the jump in depth across the bore using Eq.~\eqref{eq:Antuono_u1 u2 relation}, but with $u_2$ replaced with $u_b$ from Eq.~\eqref{eq:new antuono u1 u2 relation} to account for the oblique approach angle. This gives
\begin{equation}
    z^6-c_1z^4-c_1^2z^2+c_1^3 = 2\sec^2\theta \left[ 4c_1z^4-4\sqrt{c_1}(\alpha_2-\alpha_1 + 2c_1)z^3 + (\alpha_2-\alpha_1 + 2c_1)^2z^2\right],
    \label{eq:new antuono polynomial}
\end{equation} 
where $z = c_2/\sqrt{c_1}$ as before. Using Snell's law ($\sin \theta = \varepsilon U_b$) and Eq.~\eqref{eq:Antuono_U_b} with $u_1=0$ to write 
\begin{equation}
    \sec^2\theta = \frac{1}{1-\frac{\varepsilon^2}{2}(z^4+c_1z^2)},
    \label{eq:secant definition}
\end{equation}
and inserting into Eq.~\eqref{eq:new antuono polynomial} gives the final polynomial equation for the jump in depth across the bore. This, combined with Eq.~\eqref{eq:Antuono_U_b} and 
\begin{equation}
    \frac{dx_b}{d\tau} = U_b(h_1,u_1,\tau) \sec \theta , \quad x_b(0) = -1,
    \label{eq:new antuono _x_b}
\end{equation}
gives a closed system of equations for the obliquely approaching bore. 
    
Thus far, the obliquely approaching bore problem has been treated as a fully two-dimensional problem. We can simplify the problem by assuming a small effective approach angle (small $\varepsilon$; see \S \ref{sec:Nonlinear shallow water equations}) to derive the equations for the weakly two-dimensional case. Using Taylor series expansions of the trigonometric functions and dropping terms of $O(\varepsilon^2)$ or higher, we find that the cross-shore bore dynamics are identical to the one-dimensional case. Therefore, for the weakly two-dimensional case, there is a one-way coupling from the solution for the purely cross-shore bore dynamics (\S \ref{sec:Review of one-dimensional flow solutions}) to the alongshore flow. 

To calculate the alongshore flow just behind the bore with knowledge of the corresponding cross-shore flow, we approximate Snell's law in the small-$\theta$ limit to get
\begin{equation}
    \sin \theta = \varepsilon U_b \approx \tan \theta,
    \label{eq:tan_theta}
\end{equation}
where we have again neglected terms that are $O(\varepsilon^2)$ or higher. This allows us to write the alongshore velocity as
\begin{equation}
    v_2=\varepsilon u_2U_b.
    \label{eq:v_2 definition}
\end{equation}

To obtain the alongshore flow at the shoreline after bore collapse, we observe that the shoreline is a moving free-surface subject to the kinematic free-surface boundary condition
\begin{equation}
    u_s = \frac{\partial x_b}{\partial \tau} + v_s\frac{\partial x_b}{\partial y} \approx \frac{\partial x_b}{\partial \tau}.
    \label{eq:new antuono kfsbc}
\end{equation}
Here, $u_s$ and $v_s$ are the cross-shore and alongshore components of the shoreline velocity, respectively, for the swash and the last approximation comes from dropping a $O(\varepsilon^2)$ term. Therefore, the evolution of the shoreline angle in the swash follows
\begin{equation}
    \tan \theta = \varepsilon u_s.
    \label{eq:tan_theta_s swashzone}
\end{equation}
In fact, Eq.~\eqref{eq:tan_theta_s swashzone} is simply the small-$\theta$ approximation of Snell's law for the most forward forward-moving characteristics that describes the swash shoreline motion, and is analogous to the small-$\theta$ approximation of Snell's law for bore motion before collapse (Eq.~\eqref{eq:tan_theta}).

To calculate the full flow field, we follow the procedure outlined in \S \ref{sec:Review of one-dimensional flow solutions} for the cross-shore flow ($u(x,\tau)$ and $h(x,\tau)$). For the alongshore flow, we note that the characteristic variable related to the alongshore motion is $\gamma = (v/\varepsilon)-h-x-\tfrac{1}{2}u^2$, which is constant on curves where $dx/d\tau=u$. From the bore solution, we can calculate $\gamma_2$ and then numerically integrate the path of the $\gamma$ characteristics using knowledge of the cross-shore flow field $u(x,\tau)$. This allows us to compute $\gamma(x,\tau)$ from which we can extract $v(x,\tau)$ since $u(x,\tau)$ and $h(x,\tau)$ are already known. At the moving shoreline in the swash, since $h=0$, the characteristics for $\alpha$, $\beta$ and $\gamma$ all travel together with the shoreline, which moves at the speed $u_s$. Therefore, from Eq.~\eqref{eq:v_2 definition}, $v_s=\varepsilon U_s^2$, which shows that the alongshore component of the shoreline velocity remains constant throughout the swash. The corresponding constant $\gamma_s$ traveling with the shoreline is
\begin{equation}
    \gamma_s = \frac{1}{2}U_s^2.
    \label{eq:gamma_s}
\end{equation}

The problem is now fully solved, with both the bore (and shoreline) path and the flow behind the bore (and the shoreline) obtained. Since we referred to \citeauthor{antuono2010shock}'s cross-shore solution as the constant-$\alpha$ solution, we refer to our solution to the weakly two-dimensional system as the small-$\theta$, constant-$\alpha$ solution.

There is one subtlety related to the offshore boundary condition left to examine. We discussed in \S \ref{sec:Review of one-dimensional flow solutions} the physical interpretation of the offshore boundary condition in terms of the incoming flow (Eq.~\eqref{eq:Antuono_boundary condition}). Here, we note that there is no obvious equivalent specification for incoming alongshore flow $v^I(-1,\tau)$, since we do not specify boundary conditions for $\gamma$ at the offshore boundary. It turns out that the $\gamma$ characteristics emanating from the bore at $x=-1$ in fact initially travel onshore, resulting in a gap in the alongshore flow solution for small, positive values of $(x+1)$ and $t$, \textit{i.e.}, for small times near the offshore boundary (see Appendix \ref{sec:appA}). At later times, the problem is alleviated by offshore moving $\gamma$ characteristics arriving at $x=-1$. In lieu of a well-defined boundary condition for the incoming alongshore velocity, we postulate that the flow behind the bore maintains a constant incoming angle $\theta_0$ at the offshore boundary. From kinematics, the incoming alongshore velocity $v^I$ can then be specified as  
\begin{equation}
    v^I(-1,\tau) = u^I(-1,\tau) \tan \theta_0.
    \label{eq:incoming v}
\end{equation}
With this, we now have the information required to compute the solution throughout the domain for $x\ge -1,\tau \ge 0$. 

Finally, we note that there also exists a different analytic solution to the alongshore flow for the weakly two-dimensional system due to \citet{ryrie1983longshore}. Assuming a constant $\gamma(x,\tau_s)=\gamma_s$ throughout the swash, and using the cross-shore flow solution in Eqs.~\eqref{eq:Peregrine Williams} due to \citet{peregrine2001swash}, \citet{ryrie1983longshore} showed that the alongshore flow is given by 
\begin{equation}
    v = \frac{\varepsilon}{3}\left[ 2U_s^2 + \frac{x^2}{\tau_s^2} -U_s\tau_s + \frac{3}{4}\tau_s^2 + 2x \right]. \label{eq:Ryrie}
\end{equation}
We examine the differences between our small-$\theta$, constant-$\alpha$ solution and \citeauthor{ryrie1983longshore}'s (\citeyear{ryrie1983longshore}) constant-$\gamma$ solution (Eq.~\eqref{eq:Ryrie}) in further detail in \S \ref{sec:Numerically computed results} and \S \ref{sec:Laboratory Experiments}.

    \subsection{Numerically computed results}
    \label{sec:Numerically computed results}

Following \citet{antuono2010shock}, we compute the bore position $x_b$ from numerical integration of the bore speed (Eq.~\eqref{eq:Antuono_x_b}) using a fourth-order Runge-Kutta scheme with an adaptive time step to ensure accuracy during bore collapse given by
\begin{equation}
    \Delta\tau_2 = A[1-(1-h_1)^n], \quad \text{with} \: n=12,\: A=10^{-4}.
    \label{eq:Antuono_time stepping}
\end{equation}
The bore speed ($U_b$), the flow variables immediately behind the bore ($u_2$, $v_2$, and $h_2$), and the resulting $\beta_2$ and $\gamma_2$ values, are calculated at each time step by solving the polynomial in Eq.~\eqref{eq:Antuono_Polynomial} and using Eq.~\eqref{eq:v_2 definition}. The angle $\theta$ at each time step is calculated using Eqs.~\eqref{eq:tan_theta}, \eqref{eq:tan_theta_s swashzone}. With this procedure, the absolute error between the prescribed value of $\alpha_2$ and the value computed from numerical calculation remained less than $10^{-10}$. 

To compute the full flow field ($u(x,\tau)$, $h(x,\tau)$, $v(x,\tau)$ and $\theta(x,\tau)$), we use a grid with $\Delta x=5 \times 10^{-4}$ and $\Delta\tau=5 \times 10^{-4}$. We use linear interpolation to compute the $\beta(x,\tau)$ field after using Eq.~\eqref{eq:Antuono_beta x} at each time step of the bore solution. With this, both $\alpha$ and $\beta$ fields are known and we can compute the cross-shore flow ($u(x,\tau)$ and $h(x,\tau)$) using Eq.~\eqref{eq:Antuono_u c equation using alpha and beta}. We compute the alongshore flow ($v(x,\tau)$) from the path of each $\gamma_2$ characteristic by numerically integrating Eq.~\eqref{eq:gamma} using a fourth-order Runge Kutta scheme, using linear interpolation to obtain $\gamma(x,\tau)$, and then substituting the cross-shore flow solution into the definition of $\gamma$. 

For $x<0$, our methods yielded solutions for $h_1 \geq 10^{-12}$, after which $\Delta\tau_2$ became too small to handle numerically. %Due to numerical limitations, $c_2 \sim \textit{O}(10^{-3})$ at the end of the calculation. Since $\varepsilon\tau_2$ becomes infinitesimally small for smaller $h_1$, 
For $x\ge0$, we used the same methods with the known shoreline motion $u_s = \alpha_2 - \tau$. The resulting small discontinuity in $u_2$ and $u_s$ near $x=0$ led to numerical errors of $O(10^{-2})$ in the $\beta$ and $\gamma$ characteristics traveling close to the shoreline. Moreover, the imposition of the approximate boundary conditions for the incoming alongshore flow velocity (Eq.~\ref{eq:incoming v}) also generated errors of a similar magnitude. However, these numerical errors are confined to be very close to the shoreline and the offshore boundary, and do not introduce any additional errors to the solution away from these regions (see Appendix \ref{sec:appA}). 

    \begin{figure}
      \centerline{\includegraphics[width=1\linewidth]{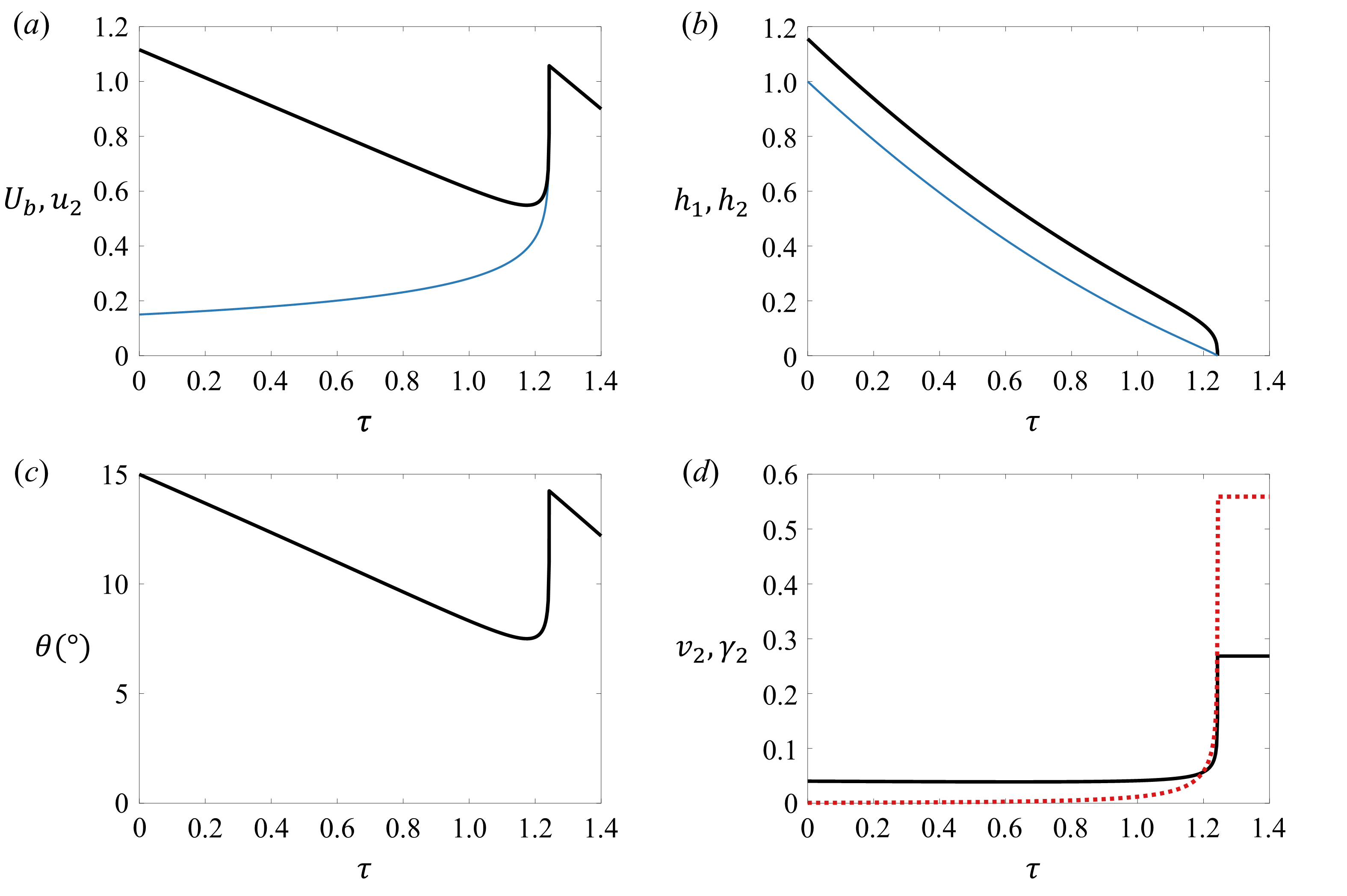}}
      \caption{Flow properties immediately behind the bore for $\alpha_2=2.3$ and $\varepsilon = 0.24$: (\textit{a}) $U_b$ (black line), $u_2$ (thin blue line); (\textit{b}) $h_2$ (black line), $h_1$ (thin blue line); (\textit{c}) $\theta$ ; (\textit{d}) $v_2$ (black line), $\gamma_2$ (red dotted line).}
    \label{fig:Flow_properties_immediately_behind_the_bore}
    \end{figure}

Figure \ref{fig:Flow_properties_immediately_behind_the_bore} shows the flow properties along the bore for $\alpha_2=2.3$ and $\varepsilon=0.24$. We chose $\alpha_2=2.3$ to reproduce the cross-shore solution in \citet{antuono2010shock}. In figure \ref{fig:Flow_properties_immediately_behind_the_bore}a, we see that $u_2$ converges with $U_b$ during bore collapse and then decreases linearly with time during the swash. In figure \ref{fig:Flow_properties_immediately_behind_the_bore}b, we see that there is a very rapid decrease in the bore height during bore collapse. The angle $\theta$ is illustrated in figure \ref{fig:Flow_properties_immediately_behind_the_bore}c, whose evolution can be understood from the small-$\theta$ approximation to Snell's law (Eq.~\eqref{eq:tan_theta}), which shows that $\tan \theta$ follows the evolution of the bore speed $U_b$ so that $\theta$ decreases during the bore's approach and then rapidly increases during bore collapse. In the swash zone, we observe a linear decrease of the shoreline angle with time, as expected from Eqs.~\eqref{eq:tan_theta_s swashzone} and \eqref{eq:Shen Meyer_u}. The alongshore velocity $v_2$ and gamma characteristics $\gamma_2$, illustrated in figure \ref{fig:Flow_properties_immediately_behind_the_bore}d, are relatively constant, but rapidly increase during bore collapse. After bore collapse, both $\gamma_2$ and $v_2$ are constant regardless of the angle $\theta$. 

    \begin{figure}
      \centerline{\includegraphics[width=1\linewidth]{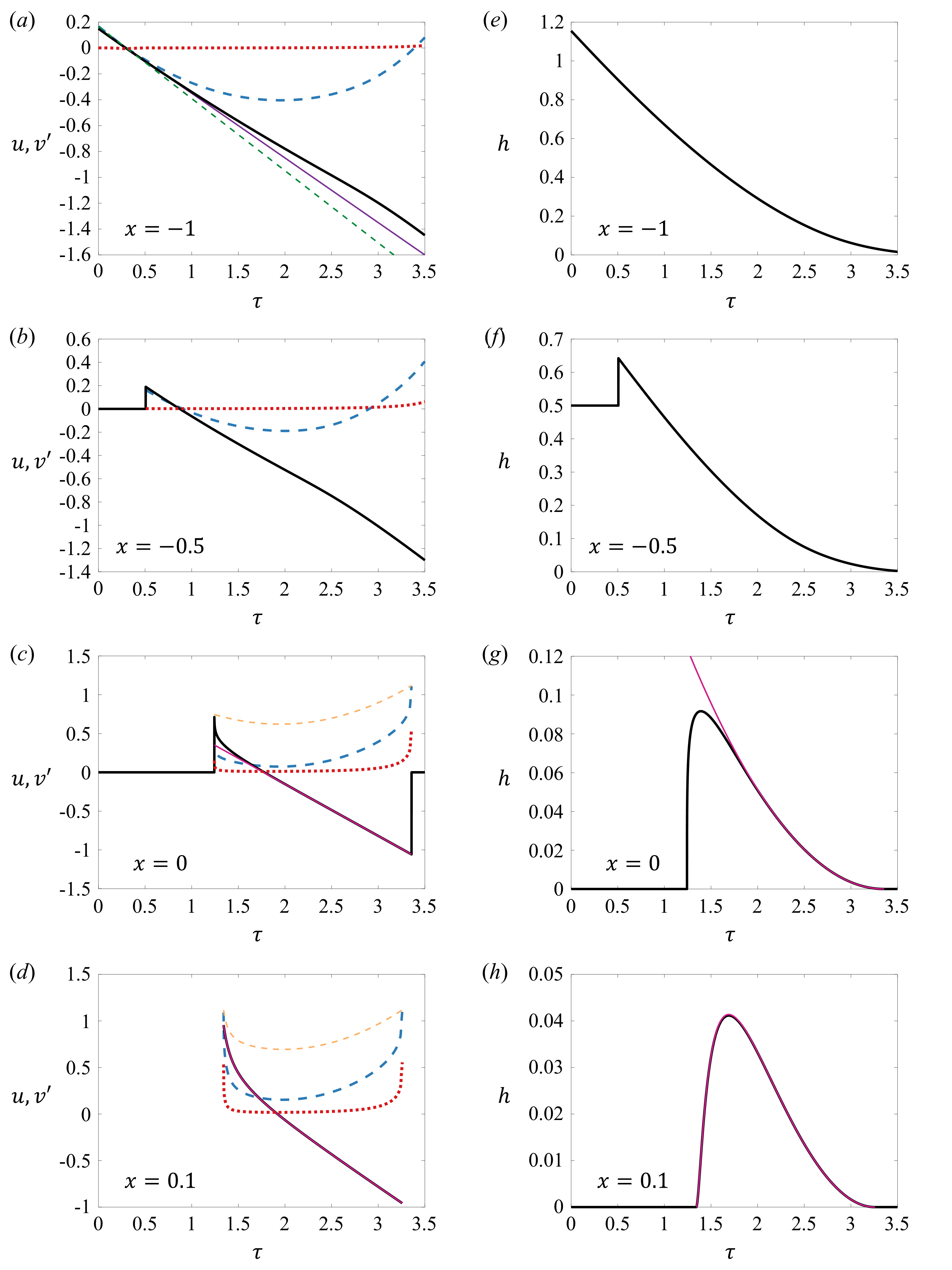}}
      \caption{Timeseries of flow velocities (\textit{a}-\textit{d}) and water depths (\textit{e}-\textit{h}) for $\alpha_2=2.3$, $\varepsilon = 0.24$. (\textit{a}-\textit{d}): $u$ (black solid line), $v' = v/\varepsilon$ (dashed blue line), $\gamma$ (dotted red line), $u^I(x=-1)$ (thin purple solid line), $v^I(x=-1)$ (thin green dashed line); (\textit{e}-\textit{h}): $h$ (black solid line); analytic solutions: $u$ and $h$ (Eqs.~\eqref{eq:Peregrine Williams}, thin magenta solid lines), $v'$ (Eq.~\eqref{eq:Ryrie}, thin orange dashed line).}
    \label{fig:Timeseries_of_u_and_h}
    \end{figure}

    \begin{figure}
      \centerline{\includegraphics[width=1\linewidth]{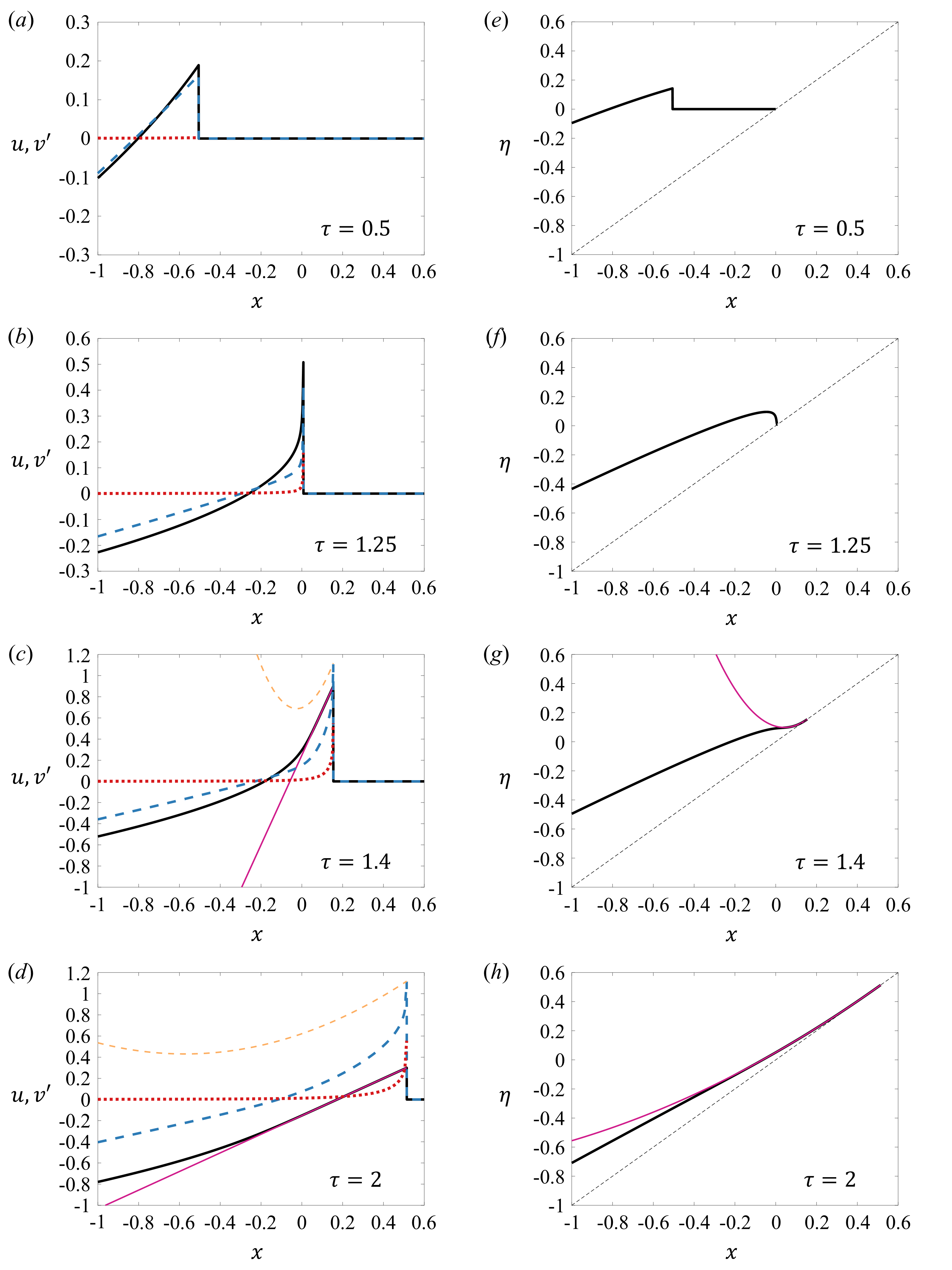}}
      \caption{Snapshots of flow velocities (\textit{a}-\textit{d}) and water depths (\textit{e}-\textit{h}) for $\alpha_2=2.3$, $\varepsilon = 0.24$. (\textit{a}-\textit{d}): $u$ (black solid line), $v' = v/\varepsilon$ (dashed blue line), $\gamma$ (dotted red line); (\textit{e}-\textit{h}): $h$ (black solid line); analytic solutions: $u$ and $h$ (Eqs.~\eqref{eq:Peregrine Williams}, thin magenta solid lines), $v'$ (Eq.~\eqref{eq:Ryrie}, thin orange dashed line).}
    \label{fig:Snapshots_of_u_and_h}
    \end{figure}

Figure \ref{fig:Timeseries_of_u_and_h} shows timeseries of the flow field at various $x$ locations. We show the alongshore velocity as $v'=v/\varepsilon$ so that its magnitude is of the same order as the cross-shore velocity and the two can be plotted together. In figure \ref{fig:Timeseries_of_u_and_h}a, the cross-shore velocity shows good agreement with its inferred value from the offshore boundary conditions ($u(-1,\tau)$, Eqs.~\eqref{eq:Antuono_boundary condition}), until the solution becomes supercritical at $\tau_{\text{crit}} \approx 1.664$ \citep{antuono2010shock}. On the other hand, the expected $v'^I(-1,\tau)$ (Eq.~\eqref{eq:incoming v}) diverges from $v'(-1,\tau)$ earlier than $\tau_{\text{crit}}$. This shows that the incoming flow does not maintain a constant angle, even before the flow becomes supercritical. 

For the swash zone ($x \geq 0$), we observe that the analytic solutions for the cross-shore flow (Eqs.~\eqref{eq:Peregrine Williams}) agree very well with our solution for larger $x$ and $\tau$. This observation is consistent with the assumptions in their derivation, which is expected to be most accurate near the shoreline. However, the analytic solution for alongshore velocity (Eq.~\eqref{eq:Ryrie}) is much larger than our small-$\theta$, constant-$\alpha$ solution. This difference stems from how bore collapse influences the alongshore dynamics, which is not taken into account in \citeauthor{ryrie1983longshore}'s analytic solution. We see from figure \ref{fig:Flow_properties_immediately_behind_the_bore}d that $\gamma_2$ and $v_2$ increase rapidly during bore collapse, which means that $\gamma$ characteristics close to the shoreline have larger values. This is reflected in figure \ref{fig:Timeseries_of_u_and_h}c-d, where we can see $\gamma$ is close to zero far from the shoreline, but rapidly increases near the shoreline. Since the analytic solution in Eq.~\eqref{eq:Ryrie} assumes that $\gamma = \gamma_s$ is constant, it predicts a larger $v$ than our small-$\theta$, constant-$\alpha$ solution.

Figure \ref{fig:Snapshots_of_u_and_h} shows snapshots of the free surface displacement and flow velocities at various times. Similar to figure \ref{fig:Timeseries_of_u_and_h}, the analytic solutions agree well with our solution close to the shoreline and for larger $\tau$ except for the alongshore velocity, which show a discrepancy. This discrepancy decreases for increasing $\tau$, but is still prominent for the reasons mentioned above. 

\subsection{Explicit expressions for minimum alongshore velocity} \label{sec:minimum_v}

    \begin{figure}
      \centerline{\includegraphics[width=1\linewidth]{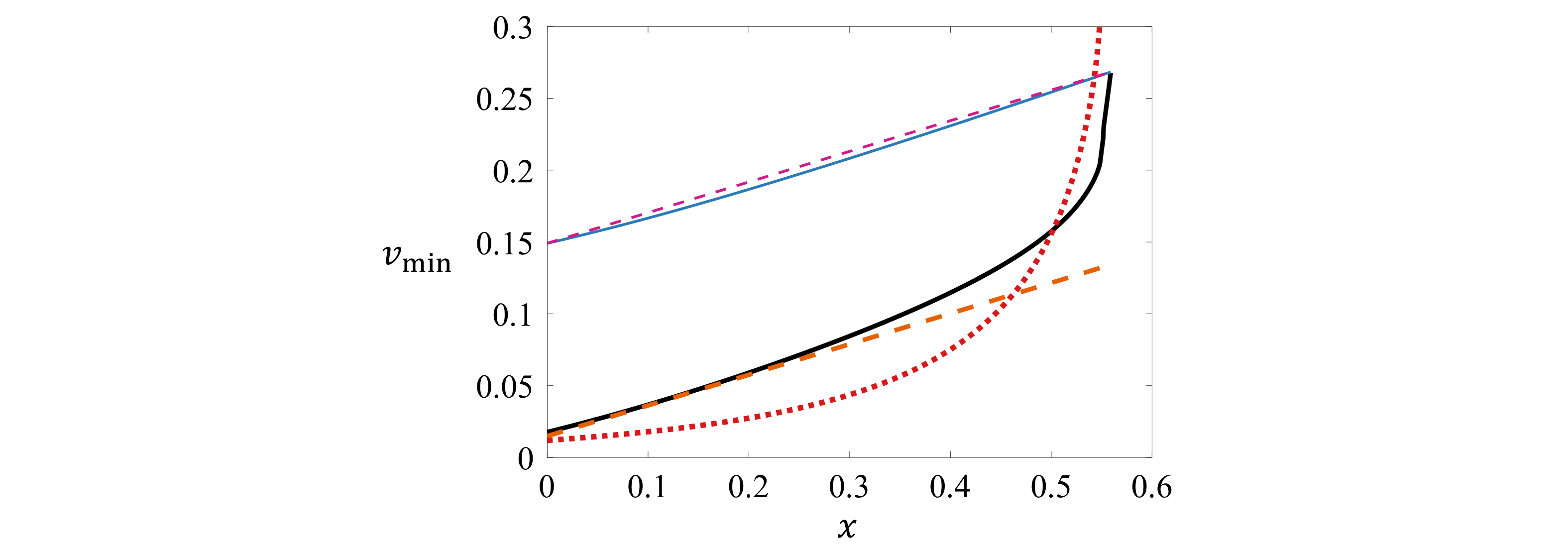}}
      \caption{Minimum alongshore velocity and $\gamma$ as a function of cross-shore position in the swash zone for $\alpha_2=2.3$ and $\varepsilon = 0.24$: $v_{\text{min}}$ for our small-$\theta$, constant-$\alpha$ solution (black solid line), $v_{\text{min}}$ for \citet{ryrie1983longshore}'s analytic solution (thin blue line),  minimum $\gamma$ for our small-$\theta$, constant-$\alpha$ solution (red dotted line), and predictions of the minimum alongshore velocities $\Tilde{v}_C$ (dashed orange line) and $\Tilde{v}_R$ (thin dashed magenta line).}
    \label{fig:Minimum_alongshore_velocity}
    \end{figure}

For predicting alongshore fluxes of fluid, solute, sediment, etc, it would be useful to have explicit formulae of the alongshore velocity as a function of cross-shore position. Since the alongshore velocity timeseries in the swash zone show that $v$ is close to its minimum value for the majority of the swash cycle, we develop predictions from the theory for this minimum alongshore velocity, $v_{\text{min}}$.

We first consider how the minimum alongshore velocity varies with cross-shore distance in the swash zone. Figure \ref{fig:Minimum_alongshore_velocity} shows $v_{\text{min}}$ for \citeauthor{ryrie1983longshore}'s analytic solution (Eq.~\eqref{eq:Ryrie}) and for the small-$\theta$, constant-$\alpha$ solution for $\alpha_2=2.3$ and $\varepsilon=0.24$. We also include data of $\gamma_{\text{min}}$ for the small-$\theta$, constant-$\alpha$ solution. While $v_{\text{min}}$ from the analytic solution shows an almost-linear increase with $x$, we observe that $\gamma_{\text{min}}$ in our solution increases with $x$ as characteristics carrying larger $\gamma$ values start traveling up the beach with higher initial velocity, resulting in a non-linear increase in $v_{\text{min}}$ with $x$ in our solution.

To obtain an explicit expression for $v_{\text{min}}$ in the swash zone for \citeauthor{ryrie1983longshore}'s analytic solution (Eq.~\eqref{eq:Ryrie}), we find the linear fit between the minimum velocity at the start of the swash zone (\textit{i.e.}, $v_{\text{min}}(x=0) = \tfrac{5}{9}\varepsilon U_s^2$) and the velocity at the maximum run-up location (\textit{i.e.}, $v(x=\tfrac{1}{2}U_s^2) = v_s$). To find an expression for $v_{\text{min}}$ for the small-$\theta$, constant-$\alpha$ solution, we note that the difference between this solution and \citeauthor{ryrie1983longshore}'s solution is the variation in the $\gamma$ value in the swash zone. In particular, \citeauthor{ryrie1983longshore}'s solution has constant $\gamma = \gamma_s$ whereas in our solution the lower swash zone (small, positive $x$) has smaller $\gamma$ values and hence smaller $v_{\text{min}}$. Given that we found that $\gamma_{\text{min}}(x\approx0) \approx 0$ for our solution (figure \ref{fig:Minimum_alongshore_velocity}), an estimate of $v_{\text{min}}$ for our solution can be obtained by subtracting $\gamma_s=\tfrac{1}{2}\varepsilon U_s^2$ from the $v_{\text{min}}$ for \citeauthor{ryrie1983longshore}'s solution. Denoting the $v_{\text{min}}$ for \citeauthor{ryrie1983longshore}'s solution as $\Tilde{v}_R$, and that for our small-$\theta$, constant-$\alpha$ solution as $\Tilde{v}_C$, we find 
\begin{subequations}\label{eq:minimum alongshore vel}
\begin{align}
  \Tilde{v}_R &= \frac{\varepsilon}{18}\left( 10 U_s^2+ 16 x\right),\label{eq:minimum alongshore vel_Ryrie} \\
  \Tilde{v}_C &= \frac{\varepsilon}{18}\left( U_s^2+ 16 x\right).\label{eq:minimum alongshore vel_Shock}
\end{align}
\end{subequations}

Figure \ref{fig:Minimum_alongshore_velocity} shows that $\Tilde{v}_R$ is very good fit to the $v_{\text{min}}$ data whereas $\Tilde{v}_C$ only captures the $v_{\text{min}}$ data accurately for small $x$ since it was derived using such an approximation. Nevertheless, both expressions provide a useful estimate of the alongshore velocity in the swash zone and show that it increases with onshore distance up to the maximum run-up location. 

\subsection{Application to laboratory or field data}\label{sec:theory_application}

    \begin{figure}
      \centerline{\includegraphics[width=1\linewidth]{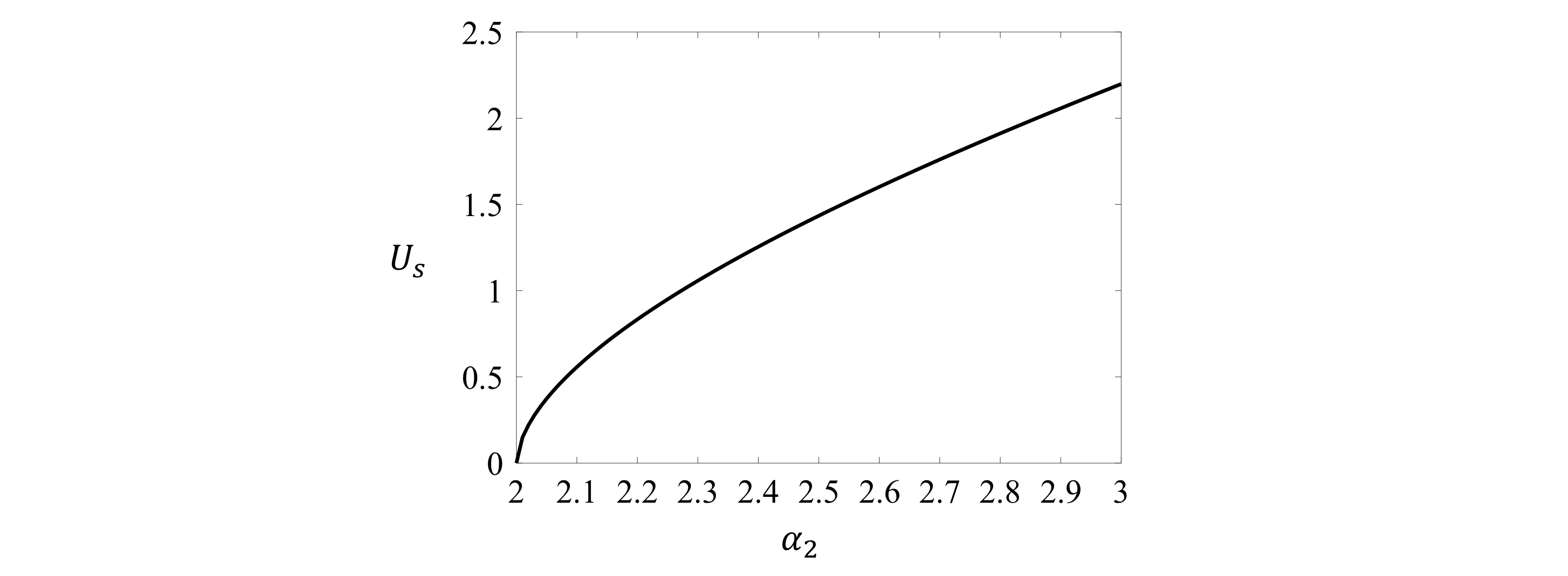}}
      \caption{The initial shoreline velocity $U_s$ with respect to $\alpha_2$.}
    \label{fig:Us_with_respect_to_alpha2}
    \end{figure}

To apply the solutions mentioned above to laboratory or field data requires several considerations. First, we must consider the real-world domain in which the NSWEs are valid and the location where we can apply the constant-$\alpha$ boundary condition. Since dispersion is neglected and bores are treated as discontinuities, the NSWEs are valid in shallow water depths where waves have already broken and the velocity field is (nearly) depth uniform. Then to apply the constant-$\alpha$ boundary condition, it becomes necessary to specify the value of $\alpha_2$, which is the only free parameter in the cross-shore solution. 

To understand the relationship between $\alpha_2$ and the properties of the incoming bore, we consider two common definitions of bore strength: \citeauthor{whitham1958propagation}'s (\citeyear{whitham1958propagation}) definition, which quantifies the degree to which the incoming bore speed is supercritical, is given in dimensionless form by $(U_{b0} - 1)$, where $U_{b0}$ is the dimensionless bore speed at the offshore boundary of the domain where the dimensionless water depth is unity. And \citeauthor{antuono2010shock}'s definition, which quantifies the magnitude of the fluid velocity immediately behind the bore, is given in dimensionless form by $(\alpha_2 - 2)$ at the offshore boundary where the dimensionless water depth is unity and at $t=0$. In both cases, it is clear that the bore strength increases with the value of $\alpha_2$.

However, as noted in \S \ref{sec:Review of one-dimensional flow solutions}, all information about the incoming bore is transferred to the initial shoreline velocity $U_s$ after bore collapse, which governs shoreline movement and is the swash zone proxy for the $\alpha_2$ value at the offshore boundary. Figure \ref{fig:Us_with_respect_to_alpha2} shows the relationship between these two variables, which is monotonic as expected. Since $U_s$ is more easily observed and measured compared to $\alpha_2$, which requires measurements of both flow velocity and water depth in the surf zone, it is easier to use data of $U_s$ and infer the implied $\alpha_2$ using figure \ref{fig:Us_with_respect_to_alpha2}.

For the alongshore component of the flow solution, the only free parameter is $\varepsilon$ since there is a one-way coupling from the cross-shore solution to the alongshore flow in the weakly two-dimensional NSWEs. Again, it is simpler to infer $\varepsilon$ from data of the shoreline motion compared to measurements of the bore during its approach to the beach. We can obtain $\varepsilon$ from measurements of the shoreline motion using the small-angle approximation of Snell's law in Eq.~\eqref{eq:tan_theta_s swashzone}.

Apart from specifying the parameters to drive the solution, there is an additional consideration when applying the constant-$\alpha$ solution to data. In figures \ref{fig:Timeseries_of_u_and_h} and \ref{fig:Snapshots_of_u_and_h}, we observe that the water depth approaches zero at all $x$ locations for large times. This occurs because the $\beta_2$ characteristics that emanate from the shoreline travel offshore without any hindrance. However, in a real-world swash flow, a hydraulic jump or backwash bore would likely form during the downrush and prevent the $\beta$ characteristics from propagating farther offshore. Thus, it is important to bear in mind that the long-time behaviour of the solution is nonphysical due to the fixed nature of the offshore boundary condition.

    \section{Laboratory experiments}
    \label{sec:Laboratory Experiments}

To examine the applicability of the theoretical solutions in \S \ref{sec:Theory}, we conducted laboratory experiments of bores approaching a beach at shore-normal and oblique angles.

    \subsection {Experimental setup}
    \label{sec:Experimental setup}

\begin{figure}
      \centerline{\includegraphics[width=0.8\linewidth]{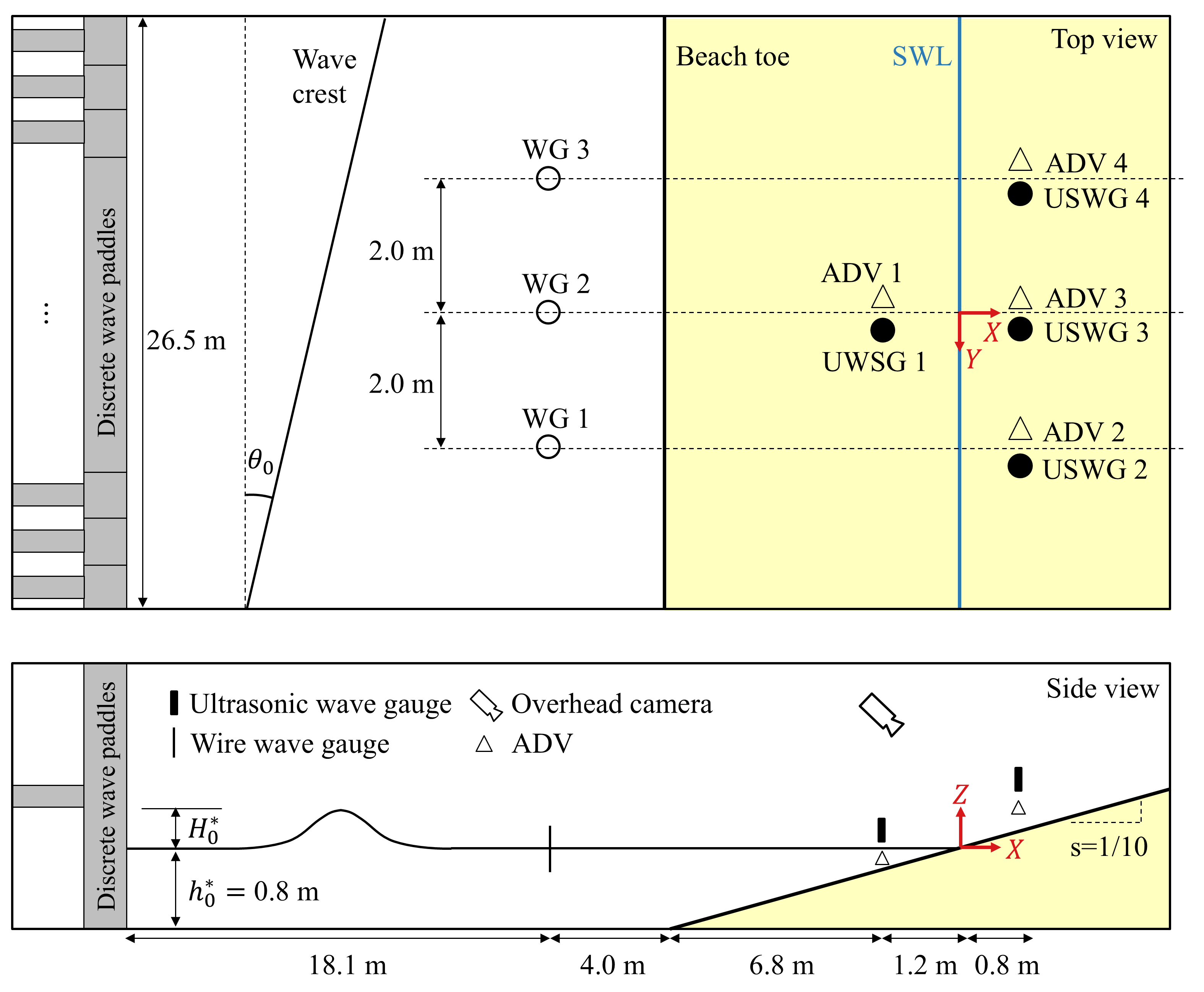}}
      \caption{Experimental setup (not to scale).}
    \label{fig:Experimental_setup}
    \end{figure}

Large-scale experiments were conducted in the Directional Wave Basin at O.H. Hinsdale Wave Research Laboratory at Oregon State University (Corvallis, OR, USA). Figure \ref{fig:Experimental_setup} shows the basin setup, where one end of the basin is equipped with a snake-type wave maker consisting of 29 discrete paddles, capable of generating multi-directional waves, and the other end has a smooth metal beach with a uniform slope of $s=1/10$. The lab coordinate system used in this study has its origin at the center of the still water line (SWL), with $X$ being the cross-shore coordinate positive onshore and $Y$ being the alongshore coordinate positive in the direction incoming oblique bores are expected to drive an alongshore flow. %\hs{Sensors not in use are not illustrated in this figure.}

The basin centreline ($Y=0$) was instrumented with one wire wave gauge (WG2, ImTech RWG) in the constant-depth region and two pairs of ultrasonic wave gauges (USWG, Senix TS-30S1) and acoustic Doppler velocimeters (ADV, Nortek Vectrino side-looking probes) at $X = -1.2$ m (USWG1, ADV1) and $X=0.8$ m (USWG3, ADV3), respectively. To quantify alongshore variability of the obliquely incident waves, two additional wire wave gauges (WG1 and WG3) were deployed at the same cross-shore location as WG2 in the constant depth region. Similarly, two additional pairs of USWG and ADV (USWG2 and ADV2, USWG4 and ADV4) were deployed at the same cross-shore location as USWG3 and ADV3. These sets of sensors were separated by $2.0$ m in the alongshore direction. Our analysis will focus on data from station 1 (S1, where USWG1 and ADV1 are located) and station 3 (S3, where USWG3 and ADV3 are located). All wave gauges and ADVs were programmed to collect data at a sampling frequency of 100 Hz and were synchronized through the data acquisition system (DAQ, National Instruments PXI-6259). To account for drift in the calibration of the wire wave gauges, we collected calibration data at the start and end of the whole experiment by slowly filling or draining the basin and recording the change in voltage. The calibration coefficients used for each experiment are the values from linear interpolation in time. 

Apart from the in situ sensors, an overhead camera with a sensor size of $1920 \times 1080$ pixels was also installed to capture images of the swash zone at a frame rate of $29.97$ Hz. These images allowed for quantifying the shoreline motion. To synchronize the camera images with the DAQ, we installed a red LED within the camera field of view, which turned on and off in an unambiguously random sequence. We recorded the voltage signal from the LED, which indicated whether it was on or off, using the DAQ and matched the LED signal from the camera with the DAQ signal to synchronize the image data with the data from the in situ sensors.

The still-water depth, which was constantly measured and monitored from a pressure sensor in the basin, was kept constant during all experiments at $h_0^*=0.80$ m. Note, as before the $*$ superscript denotes dimensional variables, but the $0$ subscript denote variables in the constant depth region for the laboratory experiments, which is different to the theory section where the $0$ subscript denoted the offshore boundary of the theoretical domain. 

To generate swash events at normal and oblique angles, we used two different methods. We generated solitary waves to obtain normally incident bores, which is a well-established method for generating isolated, but energetic swash events \citep{pujara2015swash}. However, we found in preliminary experiments that using obliquely generated solitary waves did not produce alongshore-uniform oblique bores due to their interaction with the basin side walls, in spite of the fact that the measurement section was outside of the most obvious reflection and diffraction zones near the walls. We found that obliquely propagating solitary waves undergo sufficient reflection and diffraction at the side walls that alongshore variability inevitably develops during wave travel. Under such alongshore variability, the alongshore component of the hydrostatic pressure gradient generated an unwanted alongshore flow, causing the situation to deviate from an ideal obliquely incident bore. To minimize this alongshore variability, we instead used the sidewall reflection method \citep{dalrymple1989directional, mansard1993experimental} to generate obliquely incident waves which generated oblique bores.

We used the implementation of the \citeauthor{dalrymple1989directional}'s (\citeyear{dalrymple1989directional}) method in the AwaSys wave generation software \citep[v7,][]{meinert2011awasys}. This implementation is based on an inverse diffraction calculation to obtain the target wave height and phase outside of the interference area, assuming linear superposition of the generated waves towards and away from the reflecting wall. We generated very long, regular oblique waves which were used as a proxy for solitary waves. %The time series of the paddles' positions is then modified by \citeauthor{dalrymple1989directional}'s technique. 
The wave maker trajectories from AwaSys were then used to drive the motion of the different wave paddles. By also recording the trigger signal from the wave paddle, we were able to clearly define the start time for all sensors and cameras for each wave case.

Table \ref{properties table} summarizes the properties of the 15 different wave cases used in the experiments. There were three cases of solitary waves for the zero-incident-angle tests (W13-15) and 12 cases for regular waves (W1-12) that include zero-incident-angle waves and obliquely incident waves generated using the sidewall reflection method. We generated the regular waves at two different periods, $T_0^* = 10.4$ s and $T_0^* = 8.4$ s, while also varying the wave height and incident angle. Analysis of WG2 data confirmed that the generated waves were of the desired periods.

For all cases, the incident wave height in the constant depth region, $H_0^*$, was obtained by averaging the wave heights measured at WG1–3, which is reported in table \ref{properties table}. We also report the wave breaker type according to visual observations. Specifically, $T_0^* = 8.4$ s waves displayed collapsing breakers, which exhibited weak bore collapse with a substantial amount of white foam on the bore front. We noticed that the swash period of these waves was close to the wave period, so that there may have been a very weak interaction between the swash events of successive waves that could influence their swash flows \citep{pujara2015experimental, meza2024flow}. On the other hand, the $T_0^* = 10.4$ s waves exhibited plunging breakers and it was clear that each swash event was independent so that the entire swash cycle from bore collapse to backwash bore generation were unaffected by the preceding or following wave.

Finally, we also report the nominal wave incidence angle in the constant depth region, denoted as $\theta_0$. We estimated $\theta_0$ using the measured time delay, $\Delta t^*$, between the arrival of the wave crests at WG1–3. By using $c^* = \sqrt{g^*(h_0^* + H_0^*)}$ for the phase speed, and the alongshore separation distance between WG1 and WG3, $\Delta y^* = 4\text{ m}$, we found the angle $\theta_0$ from the relation $\Delta y^* \sin \theta_0 = c^* \Delta t^*$. While this method is not exact, it provides a useful estimate of the nominal wave incidence angle.

\begin{table}
  \begin{center}
\def~{\hphantom{0}}
  \begin{tabular}{l|cccccccccccc}
      \multirow{2}*{Wave} & \multirow{2}*{$H_0^*$ (m)} & \multirow{2}*{$T_0^*$ (s)} & Breaker & \multirow{2}*{$X_c^*$ (m)} & \multirow{2}*{$\theta_0$ ($^{\circ}$)} & \multirow{2}*{$\theta_c$ ($^{\circ}$)} & \multirow{2}*{$U_{s,\alpha}(S1)$} & \multirow{2}*{$U_{s,\alpha}(S3)$} & \multirow{2}*{$U_{s,m}$} & \multirow{2}*{$\varepsilon$} \\
     &  &  & Type &  &  &  &  &  & & \\
     \hline 
    W1 & 0.182 & 10.4 & PL & -0.26 & 10.4 & 14.5 & 1.490 & 1.458 & 1.624 & 0.153\\   
    W2 & 0.172 & 10.4 & PL & -0.43 & 17.3 & 19.7 & 1.510 & 1.457 & 1.415 & 0.232\\   
    W3 & 0.214 & 10.4 & PL & -0.35 & 11.1 & 16.1 & 1.495 & 1.495 & 1.628 & 0.167\\   
    W4 & 0.201 & 10.4 & PL & -0.47 & 17.3 & 20.1 & 1.530 & 1.486 & 1.516 & 0.221\\   
    W5 & 0.239 & 10.4 & PL & -0.33 & 10.7 & 16.3 & 1.519 & 1.514 & 1.710 & 0.161\\   
    W6 & 0.228 & 10.4 & PL & -0.43 & 17.5 & 20.4 & 1.545 & 1.494 & 1.478 & 0.212\\   
    W7 & 0.216 & 8.4 & CL & -1.09 & - &- & 1.929 & 1.694 & - & -\\   
    W8 & 0.210 & 8.4 & CL & -1.06 & 5.9 & - & 1.852 & 1.633 & - & 0.088\\   
    W9 & 0.235 & 8.4 & CL & -1.13 & - & - & 1.947 & 1.730 & - & - \\   
    W10 & 0.232 & 8.4 & CL & -1.01 & 5.9 & - & 1.856 & 1.635 & - & 0.093 \\   
    W11 & 0.248 & 8.4 & CL & -1.14 & - & - & 1.921 & 1.773 & - & - \\   
    W12 & 0.248 & 8.4 & CL & -1.18 & 6.4 & - & - & 1.662 & - & 0.090 \\
    W13 & 0.220 & $\infty$ & PL & 0.53 & - & - & 1.567 & 1.475 & - & - \\
    W14 & 0.260 & $\infty$ & PL & 0.58 & - & - & 1.569 & 1.498 & - & -\\
    W15 & 0.289 & $\infty$ & PL & 0.59 & - & - & 1.588 & 1.514 & - & -\\
  \end{tabular}
  \caption{Wave properties. The subscript 0 denotes data related to the constant depth region and the subscript C denotes data related to the bore collapse location. CL are collapsing breakers and PL are plunging breakers. $T_0^*=\infty$ indicates solitary wave.}
  \label{properties table}
  \end{center}
\end{table}

    \subsection {Data processing}
    \label{sec:Data processing}

Measurements in the surf and swash zones are known to suffer from data quality issues related to bubbles, shallow water depths, and wave-to-wave variability. To improve data quality, we rely on various data processing methods for quality control. We use slightly different methods for the oblique regular waves (W1-12) and the normal solitary waves (W13-15), but in all cases, the end product is reliable and accurate data of the mean (ensemble averaged) flow. 

For W1–12, we remove obviously nonphysical spikes in the USWG data by removing data points where the magnitude of the derivative of the surface elevation timeseries exceeds 3 m/s. For the ADV data, we remove data points where the correlation coefficient (COR) is lower than 80\% and the signal-to-noise ratio (SNR) is lower than 14 dB. After that, we ensemble average the data to obtain the mean flow timeseries. In this averaging, we only use the data from the 5th to 14th waves generated by the wave paddle. The first four waves are excluded because wave gauge data clearly show signs of wave paddle start up, whereas waves after the 15th wave are excluded to avoid effects due to recirculation flows within the basin. We then identify individual swash cycles at each measurement station by detecting a sudden increase in water depth from the USWG data. By resampling each swash cycle onto a dimensionless time vector normalized by its swash duration with a resolution of 0.01 (100 data points), we obtain a total of 30 waves (10 waves $\times$ 3 repeated runs of each wave case) that are used in ensemble averaging to obtain the mean flow. 

To check whether there was significant alongshore variability in water depths and flow velocities for obliquely incident waves (W1-12), we compared data across ADV2–4 and USWG2–4. Across all wave cases, we found a maximum alongshore difference in measured velocity at any time to be smaller than 0.2 m/s (which is less than $5\%$ of the velocity scale of swash flow). Similarly, we found a maximum alongshore difference in measured water depth to be smaller than 30$\%$ of the maximum water depth at any time. These values being small, we believe the influence of the alongshore variability is minimal, and the ensemble-averaged flow field closely resembles an alongshore-uniform bore (wave height and angle being constant in $Y$) approaching a beach. 

For W13–15, we conducted 10 repetitions for each case. Unlike the regular waves, these experiments are highly repeatable since there is no wave-to-wave variability. After matching the time vectors using the wave paddle start trigger signal, we compute ensemble averages across the 10 waves (repeated experiments) and compute the mean flow. Before calculating the ensemble averages, we removed spikes in the USWG data as before. However, for the ADV data, we used a slightly different procedure to remove data points. The SNR and COR for the solitary waves were lower than for the regular waves, likely because the solitary waves did not uniformly mix the seeding tracer particles in the basin. Therefore, we first removed data points with a COR value lower than 40$\%$ and after that, we applied a modified version of the phase-space thresholding method \citep{goring2002despiking}. This iterative method, based on normal probability distribution theory, was designed for steady flow, zero-mean data (\textit{e.g.}, channel flow after subtracting the mean velocity). Since the solitary wave experiments were found to be highly repetitive \citep[see also][]{pujara2015experimental}, we removed the mean flow computed from the ensemble averaging. The algorithm was then applied iteratively, as designed. We required a minimum of five data points when calculating the ensemble-averaged velocity. If fewer than five data points were available (due to having been removed for poor quality, for example), the ensemble-averaged velocity was not calculated. Additionally, before the first iteration, obvious outliers (velocity differences larger than $0.7$ m/s and $0.2$ m/s for the cross-shore and alongshore directions, respectively) were removed while obtaining the ensemble-averaged velocity. During the iteration loops, more noisy data points were removed and the ensemble averaged velocity was updated at each iteration. The iteration process ended when no further data points were flagged for removal. The remaining velocity data were then used to compute the final ensemble average and used as the mean flow.

    \begin{figure}
      \centerline{\includegraphics[width=1\linewidth]{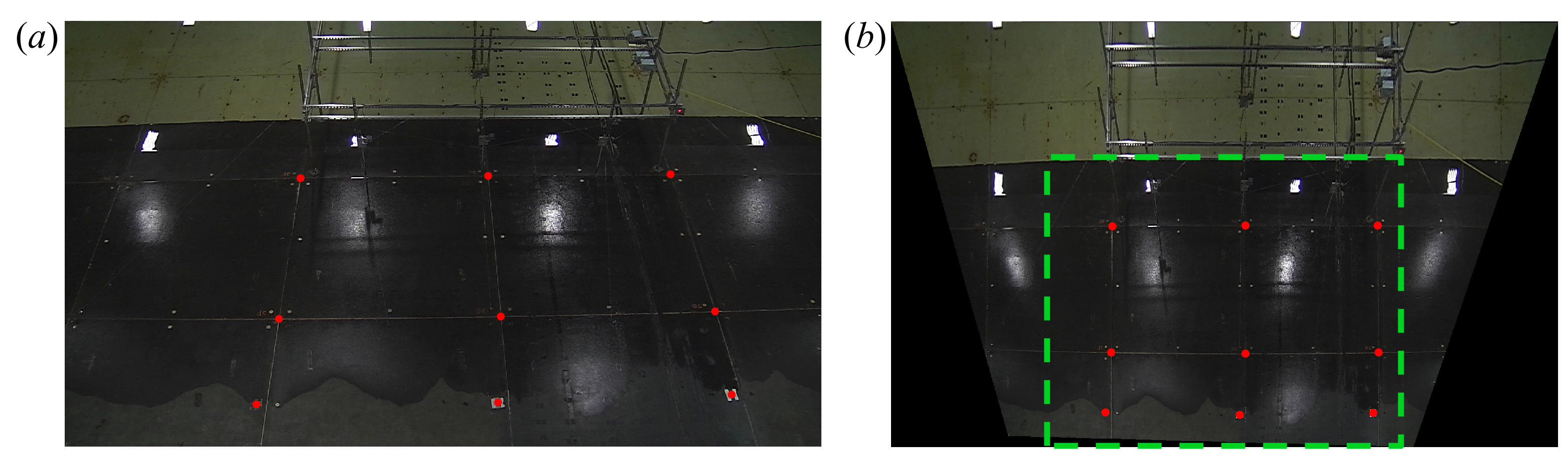}}
      \caption{Example of image rectification: (\textit{a}) Raw image, (\textit{b}) Rectified image. Red dots indicate the points used to calculate homograph transformation matrix and the green box denotes the region of interest (ROI) for calculating the shoreline movement.}
    \label{fig:Example_image_rectification}
    \end{figure}

    \begin{figure}
      \centerline{\includegraphics[width=1\linewidth]{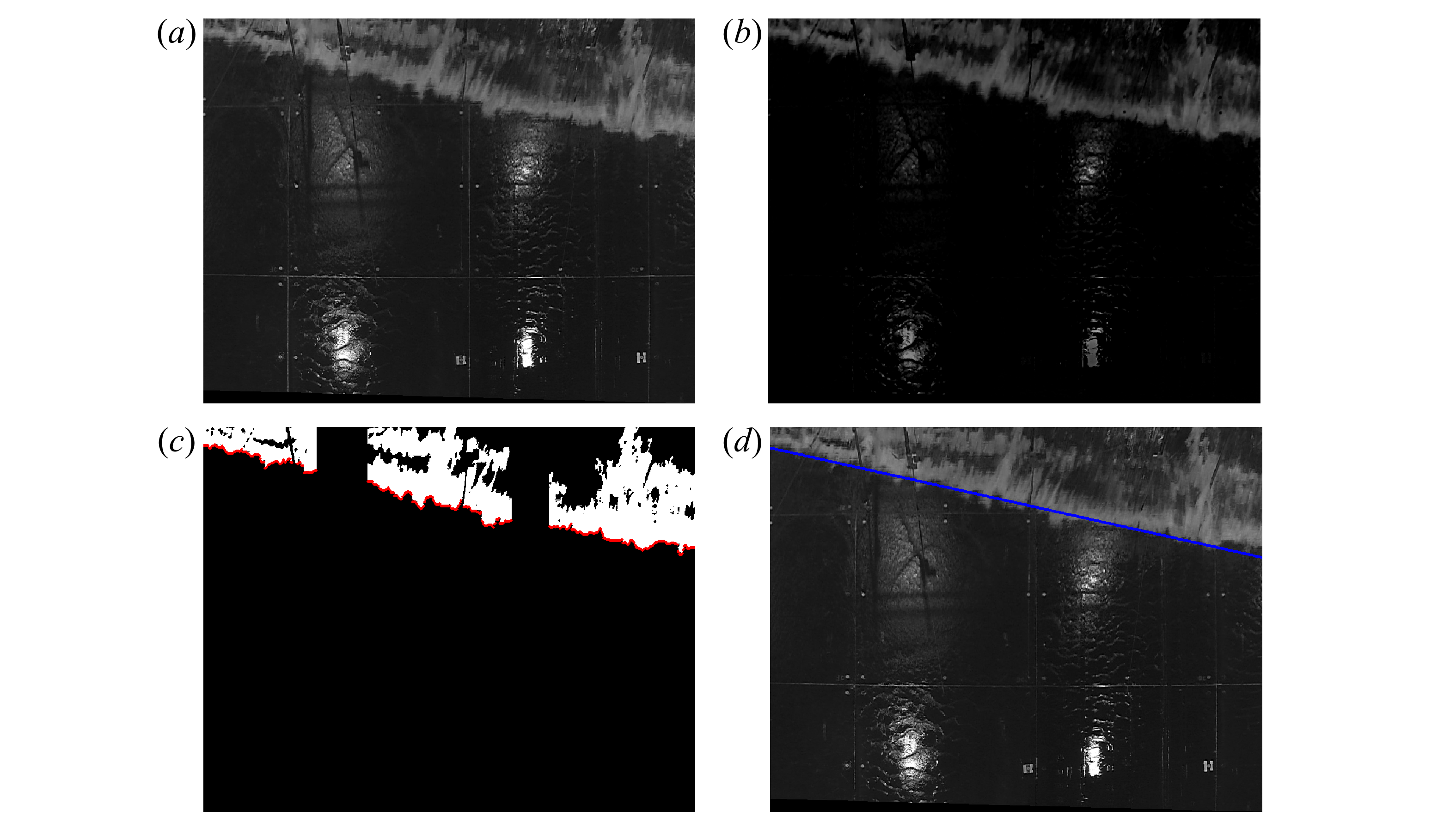}}
      \caption{Example of image processing to identify shoreline position: (\textit{a}) raw image, (\textit{b}) after background removal, (\textit{c}) after binarization with partially detected shoreline colored in red, (\textit{d}) fitted shoreline in blue on raw image.}
    \label{fig:Example_image_processing}
    \end{figure}
    
We calculated the shoreline velocity and shoreline angle using image data from the overhead camera. A single video was recorded for each wave case. Since the camera was installed obliquely to the beach, we first rectified the raw images following methods previously outlined in \citet{sung2022plastron}. Figure \ref{fig:Example_image_rectification} shows an example of a raw image and its rectified version. Briefly, using points with known lab coordinates (red dots in figure \ref{fig:Example_image_rectification}), we calculated the homography matrix which transforms the raw image into a rectified image with a resolution of 1 cm/pixel. The pixel intensities of the rectified image were calculated using bilinear interpolation. In the rectified image, we calculated the shoreline movement within the region of interest (ROI; dashed green box in figure \ref{fig:Example_image_rectification}b). The cross-shore position of the ROI was informed by the change in colour of the beach surface offshore from the selected ROI, which make it challenging to distinguish the shoreline. The alongshore extent of the ROI was chosen to be within the centre of the image where effects of lens distortion were small and between the regions where overhead lighting caused significant glare on the beach surface. 
    
Our image processing procedure to identify the shoreline position involved standard methods including background removal, binarization using a threshold, and edge detection. Figure \ref{fig:Example_image_processing} shows examples of each step. We obtained the background image as the mean across all images of a given video. By subtracting this background, we were able to enhance the contrast between the beach and the approaching shoreline (figure \ref{fig:Example_image_processing}b). We used a normalised binarization threshold of $0.09$, and then removed regions affected by glare, by structures holding the sensors, and regions smaller than 300 px in area to give the final binarized image to be used for shoreline detection (figure \ref{fig:Example_image_processing}c). We detected the shoreline edge by searching for the lowest (most onshore) remaining pixel in each column (shown in red in figure \ref{fig:Example_image_processing}c). Finally, we fit a straight line to the edge detected points to find the shoreline position and angle (shown in blue in figure \ref{fig:Example_image_processing}d).

    \subsection {Properties of swash flow}
    \label{sec:Properties of swash flow}

\subsubsection{Bore collapse and shoreline motion}

    \begin{figure}
      \centerline{\includegraphics[width=1\linewidth]{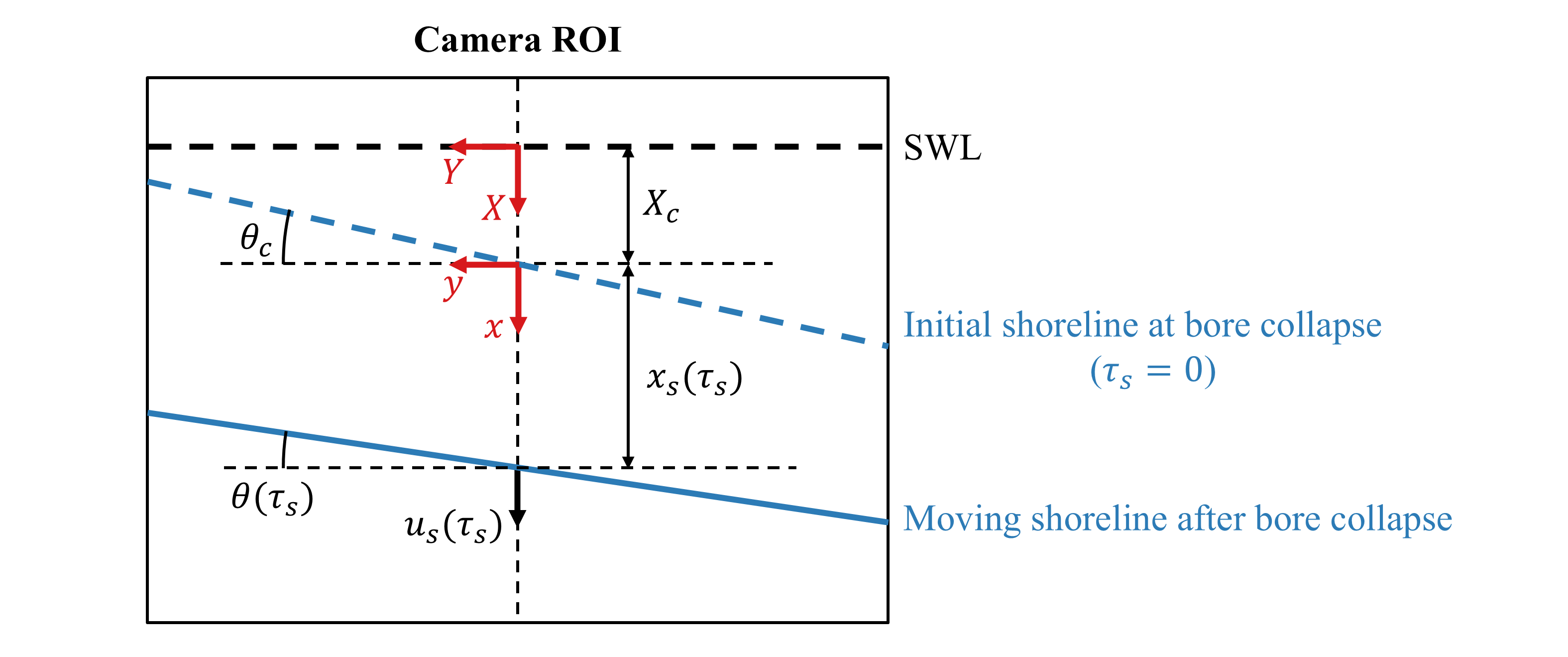}}
      \caption{Diagram to illustrate shoreline properties.}
    \label{fig:Diagram_shoreline_properties}
    \end{figure}

Bore collapse is an important feature of the experimental data since it defines the start of the swash. Figure \ref{fig:Diagram_shoreline_properties} illustrates the different laboratory coordinate systems and variables related to the bore collapse process in the experiments. We define the coordinate system relative to the still water line as $(X,Y)$ (see figure \ref{fig:Experimental_setup}) and the coordinate system relative to the bore collapse location as $(x,y)$. The distance between the origins of these two locations, denoted $X_c$, quantifies where bore collapse occurs relative to the still water line, and $\theta_c$ quantifies the bore angle at collapse. Table \ref{properties table} lists these values, which were measured by manually capturing the bore collapse at the basin centerline ($Y=0$) in the camera images and reported as the ensemble mean values over 10 waves. The time of bore collapse is also used to define the swash zone pseudo-time origin, $\tau_s = 0$.

It is interesting to note that the solitary waves collapse onshore of the still water line ($X_c>0$) since the fluid velocity is always pointed onshore, whereas the regular waves collapse offshore of the still water line ($X_c<0$) due to the presence of a wave trough and the associated offshore directed fluid velocity. Note, the collapse location was too far offshore of the camera ROI to be able to obtain a reliable measurement of the bore angle at collapse for the $T_0^* = 8.4$ s waves.
       
    \begin{figure}
      \centerline{\includegraphics[width=1\linewidth]{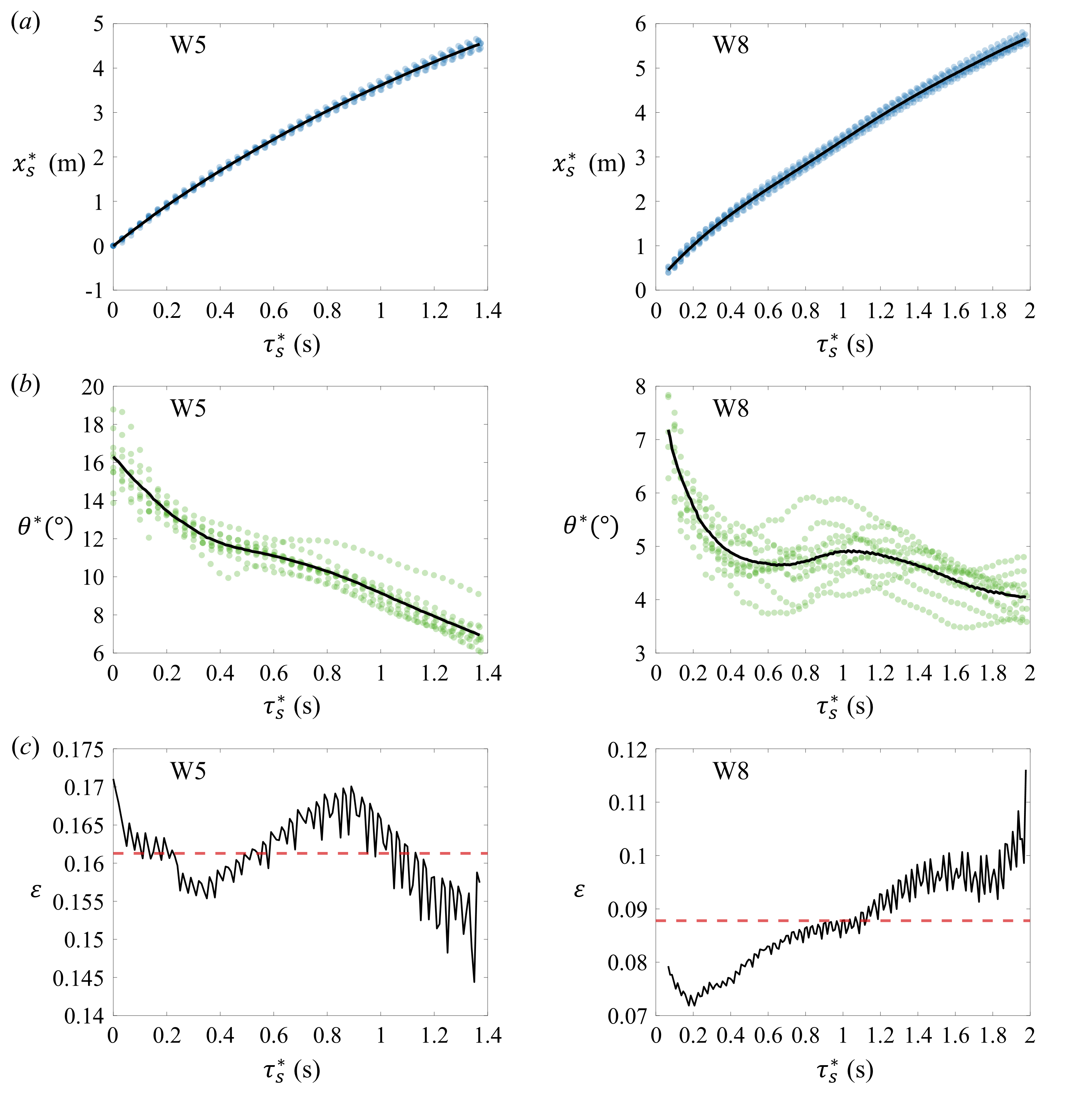}}
      \caption{Shoreline motion data for W5 and W8:(\textit{a}) shoreline position $x_s^*$, (\textit{b}) angle $\theta$, (\textit{c}) effective bore approach angle $\varepsilon$. In (\textit{a},\textit{b}): dots show raw data from 10 individual waves and solid lines show ensemble average. In (\textit{c}), solid line shows the time evolution and red dashed line shows the time mean.}
    \label{fig:Shoreline_motion_data}
    \end{figure}  

We track the shoreline motion after bore collapse, and extract the timeseries of shoreline position $x_s$ and angle $\theta$. Figure \ref{fig:Shoreline_motion_data}a,b show example timeseries of this data for W5 and W8, where raw data from individual waves are shown as dots and the ensemble averaged values as solid lines. To reduce noise and allow for calculating the shoreline velocity in a single step, we compute the ensemble averages using linear fits over a moving window where the window size (typically $0.13$–$0.32$ s) was selected to minimize the mean variance of the difference between the raw and ensemble average data.

\subsubsection{Effective bore approach angle, $\varepsilon$}

The shoreline motion data allows us to extract the effective bore approach angle, $\varepsilon$. We do this by calculating $\varepsilon$ from the equation of shoreline angle in the swash zone under small-angle approximation (Eq.~\eqref{eq:tan_theta_s swashzone}). Figure \ref{fig:Shoreline_motion_data}c shows example results of $\varepsilon$ for W5 and W8. We observe that, overall, $\varepsilon$ is constant during the shoreline's motion during uprush. We use the time mean (shown as a dashed red line) as our estimate of the constant value, which is then used to calculate the theoretical solution. This time mean $\varepsilon$ value is reported in table \ref{properties table} and we observe that it is small ($<0.25$) across all wave cases, justifying our small-$\theta$ approximation. The relative standard deviations of $\varepsilon$ (RSD; standard deviation divided by the mean), which quantifies the magnitude of the deviations away from the mean, are $<15\%$ (collapsing breakers) and $<7\%$ (plunging breakers), showing that the shoreline motion obeys $\varepsilon \approx \text{constant}$ to a good approximation.

The shoreline motion data allows us to extract the effective bore approach angle, $\varepsilon$. We do this by calculating $\varepsilon$ from the equation of shoreline angle in the swash zone under small-angle approximation (Eq.~\eqref{eq:tan_theta_s swashzone}). Figure \ref{fig:Shoreline_motion_data}c shows example results of $\varepsilon$ for W5 and W8. We observe that, overall, $\varepsilon$ is a constant during the shoreline's motion during uprush. We use the time mean (shown as a dashed red line) as our estimate of the constant value, which is then used to calculate the theoretical solution. This time mean $\varepsilon$ value is reported in table \ref{properties table} and we observe that it is small ($<0.25$) across all wave cases, justifying our small-$\theta$ approximation. The relative standard deviations of $\varepsilon$ (RSD; standard deviation divided by the mean), which quantifies the magnitude of the deviations away from the mean, are $<15\%$ (collapsing breakers) and $<7\%$ (plunging breakers), showing that the shoreline motion obeys $\varepsilon \approx \text{constant}$ to a good approximation.

    \subsubsection{Effective initial shoreline velocity, $U_s$}
    \label{sec:Effective shoreline velocity}

\begin{figure}
      \centerline{\includegraphics[width=1\linewidth]{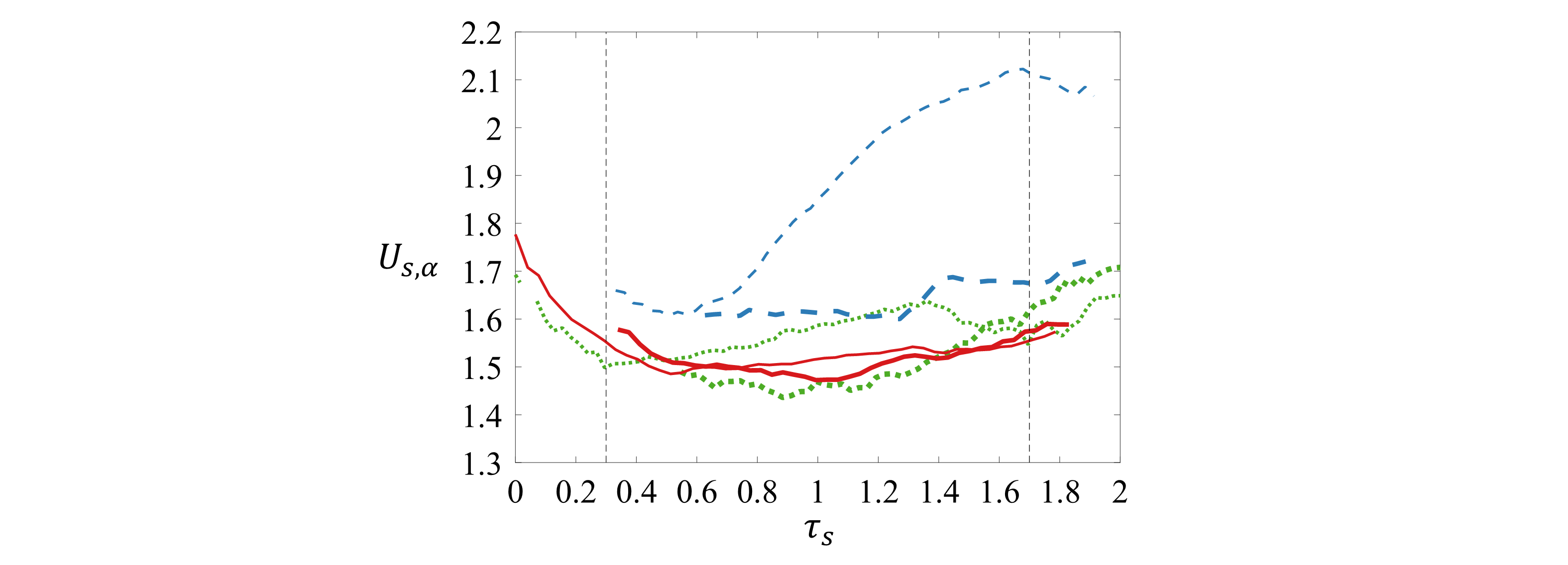}}
      \caption{Effective initial shoreline velocity from in situ sensor data at different measurement stations: $U_{s,\alpha}(S1)$ (thin line), $U_{s,\alpha}(S3)$ (thick line) for W5 (red solid line), W8 (blue dashed line), and W14 (green dotted line). Vertical dashed lines indicate a window for calculating time mean.}
    \label{fig:Effective_initial_shoreline_velocity}
    \end{figure}

As noted in \S \ref{sec:theory_application}, we can use the initial shoreline velocity $U_s$ to infer the $\alpha_2$ of the incoming bore. Following \citet{pujara2015swash}, we estimate the effective $U_s$ for the swash flow in multiple ways. 

The first is the measured initial shoreline velocity from camera data, $U_{s,m}$. We extract this from the slope of the shoreline position (figure \ref{fig:Shoreline_motion_data}a) at $\tau_s = 0$ and report it in
table \ref{properties table}. However, we were only able to extract this for the $T_0^* = 10.4$ s waves. For solitary waves, it was not possible to obtain $U_{s,m}$ due to insufficient time resolution in the camera data and the extra noise due to vibrations from the plunging breaker that shook the camera. For the $T_0^* = 8.4$ s waves, the bore collapse location was offshore of the camera ROI as explained above.

The second is to use flow data from the in situ sensors combined with the definition of the forward moving characteristic, which gives $U_{s,\alpha} = u+2c+\tau_s$. We calculate $U_{s,\alpha}(S1)$ and $U_{s,\alpha}(S3)$, using the two different cross-shore measurement stations S1 and S3, respectively. 

Figure \ref{fig:Effective_initial_shoreline_velocity} shows example timeseries of $U_{s,\alpha}(S1)$ and $U_{s,\alpha}(S3)$ for W5, W8, and W14. Other wave cases show behaviour that is very similar to the corresponding breaker type in figure \ref{fig:Effective_initial_shoreline_velocity}. We observe that, overall, the $U_{s,\alpha}$ values remain constant. This confirms the applicability of the theory since it implies constant-$\alpha$ in the swash flow. The notable exception is $U_{s,\alpha}(S1)$ for W8, which we discuss further below. For all wave cases, we calculate the time mean $U_{s,\alpha}(S1)$ and $U_{s,\alpha}(S3)$ over the range $0.3 \leq \tau_s \leq 1.7$ (dashed vertical lines in figure \ref{fig:Effective_initial_shoreline_velocity}) and use it to calculate the theoretical solution. These values are reported in table \ref{properties table}. The RSD values for $U_{s,\alpha}$ are $<2.5\%$ across all wave cases. This excludes $U_{s,\alpha}(S1)$ for collapsing breakers, such as $U_{s,\alpha}(S1)$ for W8 in figure \ref{fig:Effective_initial_shoreline_velocity}, for which the deviation is higher. 

$U_{s,\alpha}(S1)$ is not constant for collapsing breakers, but rather increases during the swash cycle. This behavior suggests that the NSWEs and constant-$\alpha$ solution may not accurately represent the flow at $x<0$ for collapsing breakers. One reason for the increasing $U_{s,\alpha}$ at locations offshore of bore collapse could be due to incomplete bore collapse. \citet{jensen2003experimental} showed with particle image velocimetry data that the velocity field is not depth-uniform for surging breakers even when the bore is in very shallow depths. A depth-dependent velocity field indicates that dispersion effects are non-negligible, which in turn means the NSWEs are not applicable. Another possible reason could be the weak wave-swash interactions observed for collapsing breakers, which may have caused the incoming waves to break earlier than expected due to offshore directed momentum from the backwash. Additionally, the mass flux from the backwash of the previous wave might contribute to an increased water depth, leading to the observed increase in $U_{s,\alpha}(S1)$ over time. Lastly, our analysis of the NSWEs ignores the effects of bed friction, which is expected to decrease the rate of water drainage during backwash \citep{pedersen2013runup} leading to larger water depths and therefore increase $U_{s,\alpha}(S1)$. However, we note that while $U_{s,\alpha}(S1)$ is not constant for collapsing breakers, $U_{s,\alpha}(S3)$ does remains relatively constant, indicating that the constant-$\alpha$ solution should apply to their swash flow.

Briefly, we also mention that $U_{s,\alpha}(S1) \ge U_{s,\alpha}(S3)$ for all cases where the data is available. Since $U_{s,\alpha}$ quantifies the net energy in the flow, we attribute the decrease to losses due to friction and turbulence. The rate at which this loss occurs, and how it depends on breaker type, is left for future study. 

To calculate the theoretical solution, we use the mean of $U_{s,\alpha}$ across time and across stations $S1$ and $S3$ for plunging breakers, and the mean $U_{s,\alpha}$ across time at station $S3$ for collapsing breakers.

    \section{Comparison of laboratory data with theory}
    \label{sec:Comparison with theoretical solutions}

We now compare the theoretical solutions (\S \ref{sec:Theory}) with the velocity and water depth data from the experiments (\S \ref{sec:Laboratory Experiments}). As noted in \S \ref{sec:theory_application}, the two parameters required for generating the weakly two-dimensional small-$\theta$, constant-$\alpha$ solution are the effective shoreline velocity $U_s$ and the effective bore approach angle $\varepsilon$. We derive the values of these parameters from the data (\S \ref{sec:Properties of swash flow}), effectively fitting the theoretical solution to the data.

    \begin{figure}
      \centerline{\includegraphics[width=1\linewidth]{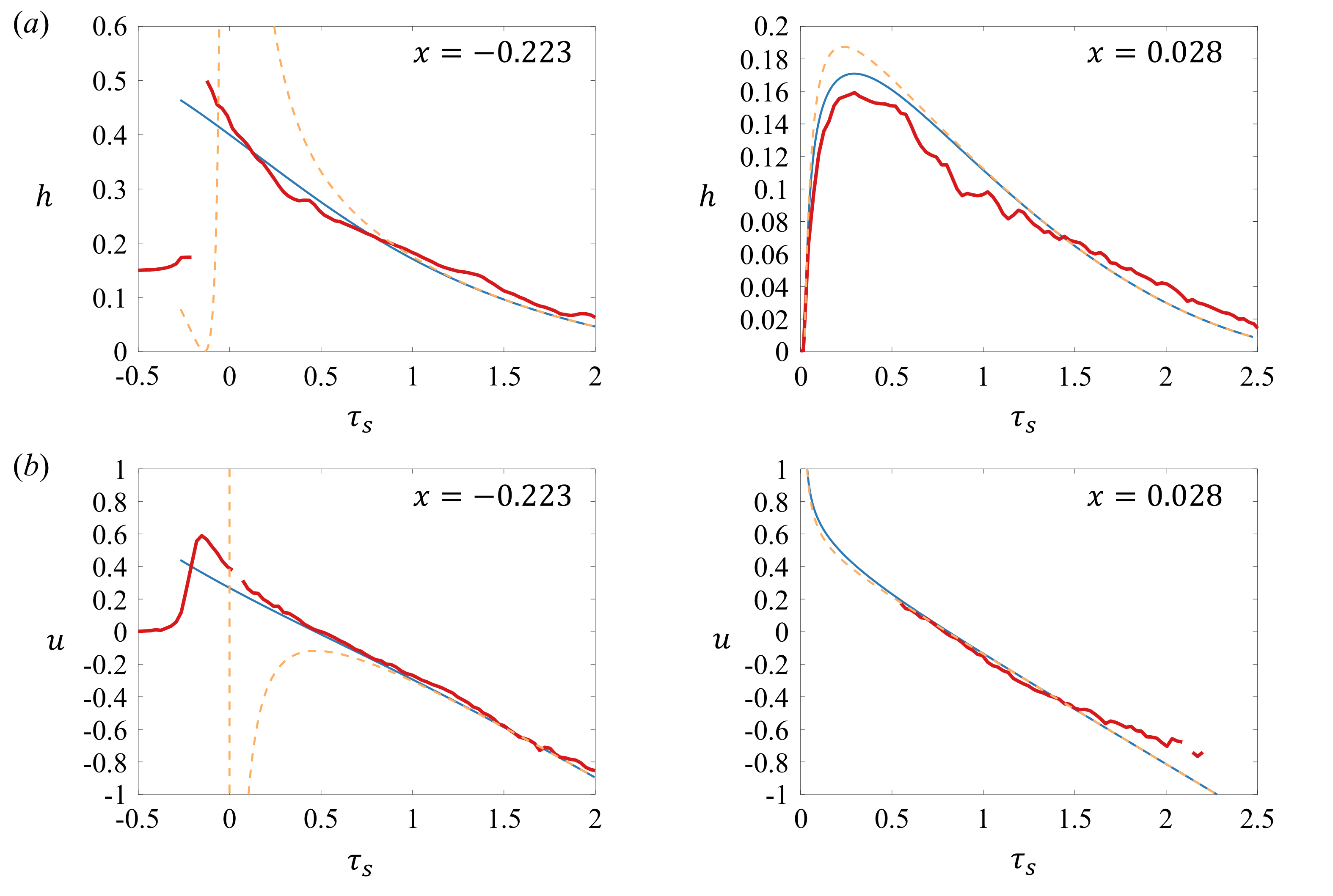}}
      \caption{Comparison between the experimental data (thick red solid line), the \citeauthor{antuono2010shock}'s (\citeyear{antuono2010shock}) constant-$\alpha$ solution (blue solid line), and the analytic solution (orange dashed line) at two different cross-shore locations for W14: (\textit{a}) water depth $h$, (\textit{b}) cross-shore velocity $u$.}
    \label{fig:Model_comparison_W14}
    \end{figure}
    
We begin the theory-data comparison with the normally incident solitary wave case. Figure \ref{fig:Model_comparison_W14} shows the comparison for W14, which includes both \citeauthor{peregrine2001swash}'s (\citeyear{peregrine2001swash}) analytic solution and \citeauthor{antuono2010shock}'s constant-$\alpha$ solution. Overall, there is good agreement between theory and data. More specifically, the analytic solution and the constant-$\alpha$ solution agree well with the data for $\tau_s \gtrsim 1$ at $x=-0.223$, but the constant-$\alpha$ solution better captures the measured flow field during the initial swash stage for $\tau_s \lesssim 1$. This is because the analytic solution is derived from expanding \citeauthor{shen1963climb}'s (\citeyear{shen1963climb}) solution, which is only asymptotically valid near the shoreline. At $x=0.028$, both theoretical solutions work well, and the discrepancy during the later stages of the swash cycle ($\tau_s \gtrsim 1.5$) can likely be attributed to slower drainage of water during backwash due to friction. We note that while good agreement between the analytic solution and data for swash flow due to normally incident solitary waves was previously shown \citep{pujara2015swash}, the ability of \citeauthor{antuono2010shock}'s constant-$\alpha$ solution to capture the flow evolution at locations offshore of the bore collapse location is new. 

    \begin{figure}
      \centerline{\includegraphics[width=1\linewidth]{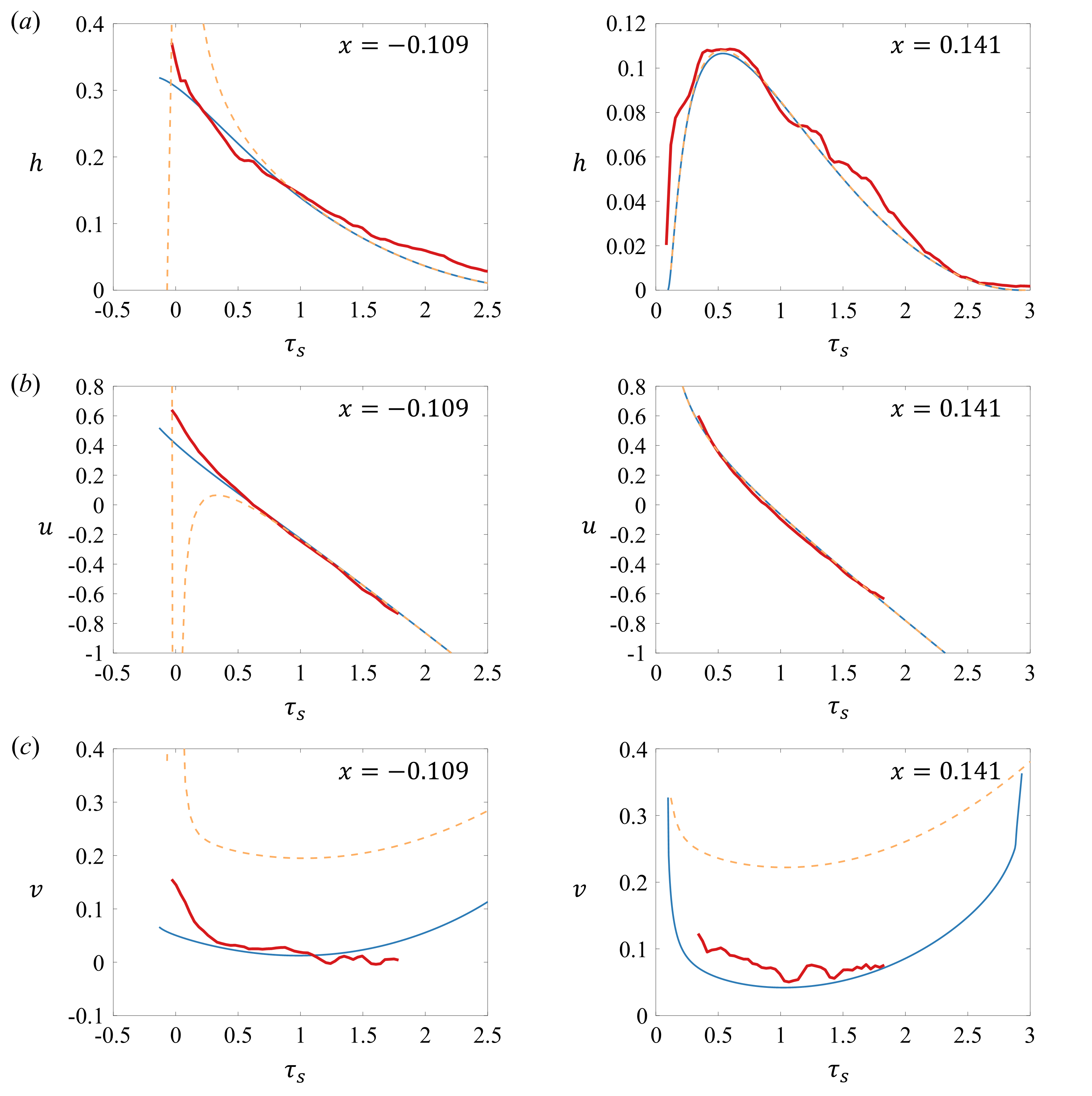}}
      \caption{Comparison between the experimental data (thick red solid line), the small-$\theta$, constant-$\alpha$ solution (blue solid line), and the analytic solution (orange dashed line) at two different cross-shore locations for W5: (\textit{a}) water depth $h$, (\textit{b}) cross-shore velocity $u$, (\textit{c}) alongshore velocity $v$.}
    \label{fig:Model_comparison_W5}
    \end{figure}
    
Next, we consider the obliquely incident plunging breakers where the nominal approach angle varied between 10$^\circ$-20$^\circ$. Figure \ref{fig:Model_comparison_W5} shows the comparison for W5 as an example of these wave cases. Here, we also show the analytic solutions by \citet{peregrine2001swash} and \citet{ryrie1983longshore} and our small-$\theta$, constant-$\alpha$ solution (\S \ref{sec:Weakly two-dimensional flow solution}). There is good agreement between theory and data in the cross-shore flow and water depth, as before. In the alongshore flow, we observe that the small-$\theta$, constant-$\alpha$ solution agrees quite well with the data and that the analytic solution does not. This is likely due to the source of the $\gamma$ characteristics that enter the swash, and in particular, it shows that $\gamma \neq \text{constant}$ in the swash zone. Due to limitations of the ADVs, we do not capture the velocity at the start and end of the swash cycles, so it is not possible to confirm that there is a rapid decrease of the alongshore velocity at the start of uprush and a rapid increase of the alongshore velocity at the end of backwash. Moreover, since frictional effects become significant when the water depth is shallow, resolving the alongshore velocity during these early and late stages of the swash cycle remains an important objective for future investigation.

    \begin{figure}
      \centerline{\includegraphics[width=1\linewidth]{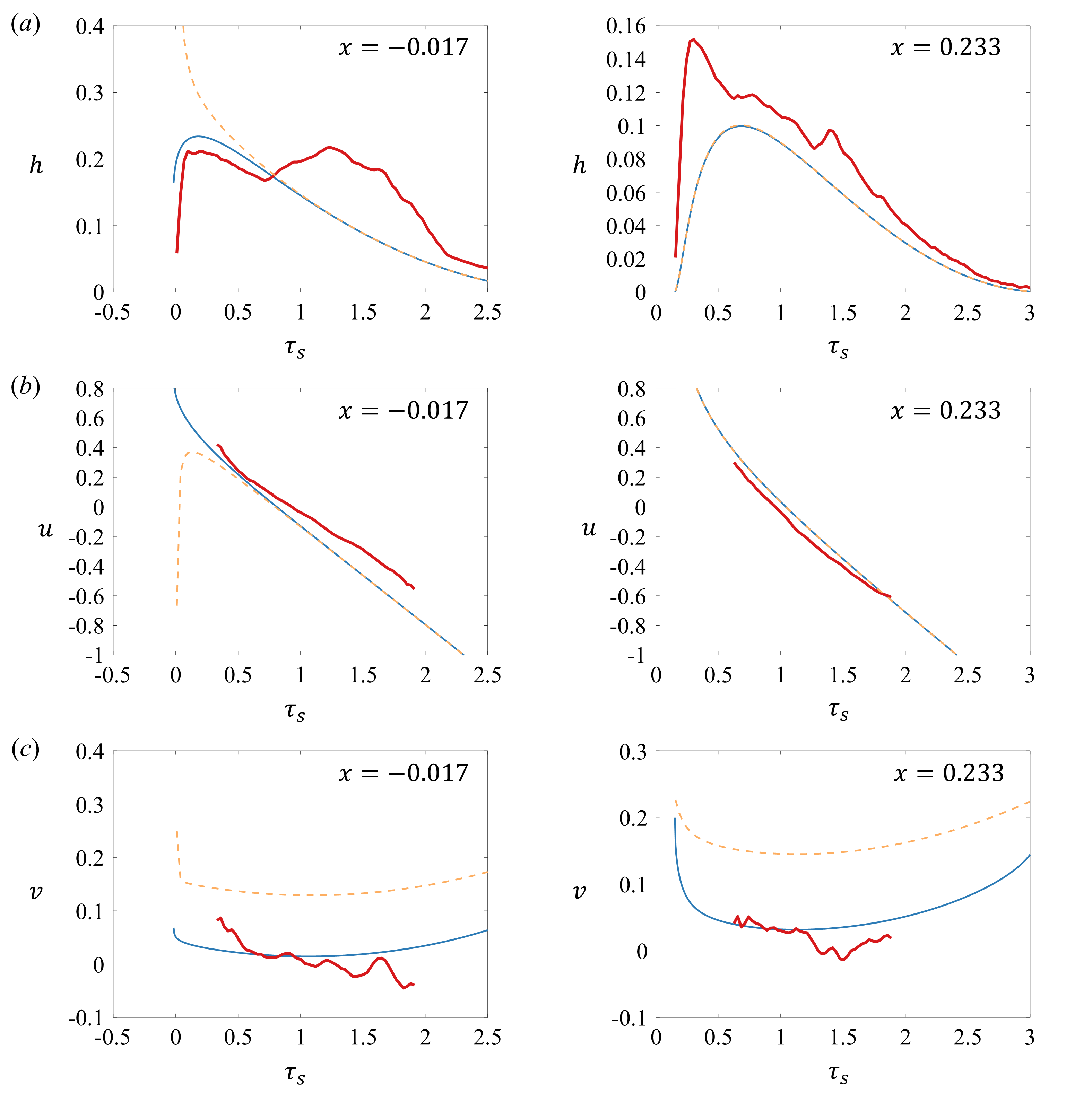}}
      \caption{Comparison between the experimental data (thick red solid line), the small-$\theta$, constant-$\alpha$ solution (blue solid line), and the analytic solution (orange dashed line) at two different cross-shore locations for W8: (\textit{a}) water depth $h$, (\textit{b}) cross-shore velocity $u$, (\textit{c}) alongshore velocity $v$.}
    \label{fig:Model_comparison_W8}
    \end{figure}

We next consider the obliquely incident collapsing breakers where the nominal approach angle varied between 0$^\circ$-10$^\circ$. Figure \ref{fig:Model_comparison_W8} shows the comparison for W8 as an example of these wave cases. As before, we also show the analytic solutions by \citet{peregrine2001swash} and \citet{ryrie1983longshore} and our small-$\theta$, constant-$\alpha$ solution (\S \ref{sec:Weakly two-dimensional flow solution}). Given that we observed that $U_{s,\alpha}(S1)$ was not a constant through the swash cycle, we do not expect good agreement between the theory and data at $x=-0.017$. However, the comparison is somewhat reasonable during the uprush portion of the swash cycle before the data starts to deviate from the theory significantly. The second increase in $h$ at $\tau_s \approx 1$ is particularly notable and might result from the development of a secondary bore and its collapse, although this is not clearly discernible from the camera data. At $x=0.233$, the agreement between the theory and data is better, but the water depth data is significantly larger than the prediction from theory. We suspect this discrepancy might be due to a bias in the USWG, which would mistake bubbly foam as water as long as it reflects sufficient sound. The camera data showed collapsing breakers produced a significant number of bubbles at the shoreline tip. Additionally, uncertainties in the bore collapse location, which would change the origins of the $(x, \tau_s)$ coordinates could also affect the comparison. Importantly, magnitude of the alongshore velocity during the middle of the swash cycle is predicted well.

     \begin{figure}
      \centerline{\includegraphics[width=1\linewidth]{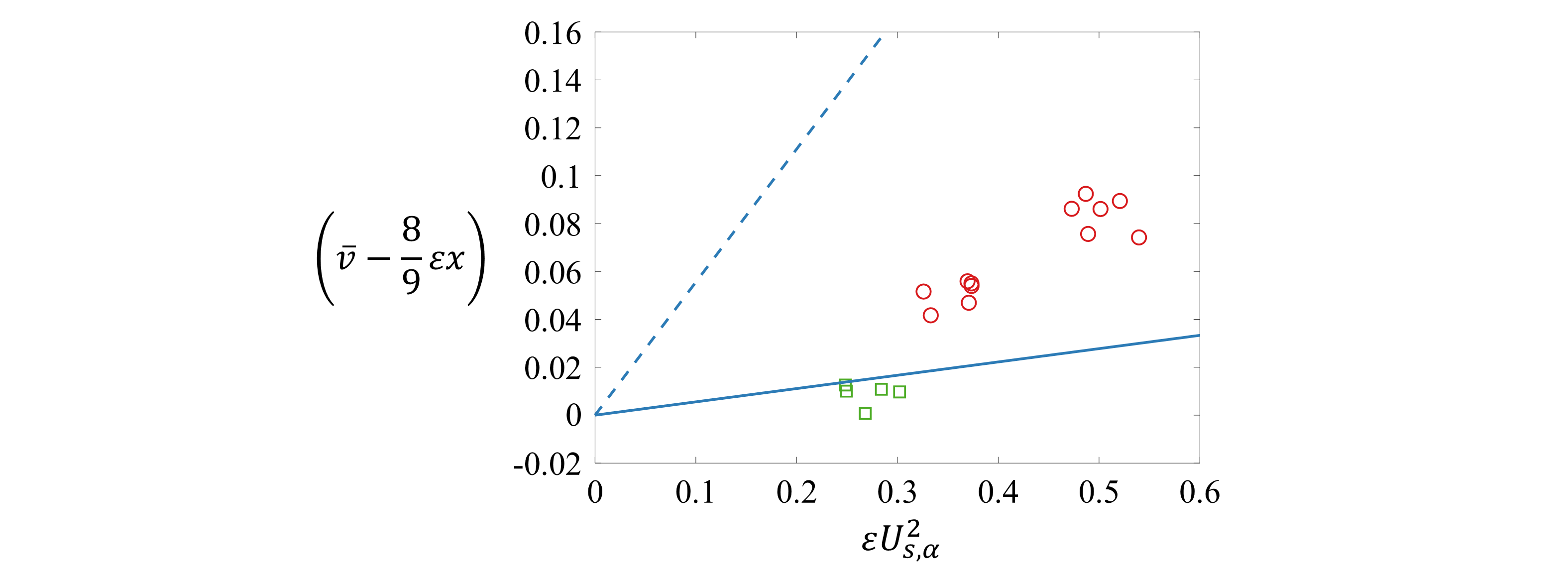}}
      \caption{The time-mean, cross-shore-distance-compensated alongshore velocity. Comparison between the experimental data and the prediction from small-$\theta$, constant-$\alpha$ solution $\Tilde{v}_C$ (blue solid line), the prediction from \citeauthor{ryrie1983longshore}'s analytic solution
      $\Tilde{v}_R$ (blue dashed line) for plunging breakers (red circles) and collapsing breakers (green squares).}
    \label{fig:Time-mean_x-compensated_alongshore_velocity}
    \end{figure} 

While figures \ref{fig:Model_comparison_W14}-\ref{fig:Model_comparison_W8} show data from representative examples, we note that other wave cases that are not shown give similar results, corresponding to the breaker type.

Finally, we compare the time-averaged alongshore velocity $\overline{v}$ over the swash cycle with the expressions for the approximate minimum alongshore velocity derived above (\S \ref{sec:minimum_v}, Eq.~\eqref{eq:minimum alongshore vel}). These are $\Tilde{v}_R$ and $\Tilde{v}_C$, respectively, for \citeauthor{ryrie1983longshore}'s analytic solution and our small-$\theta$, constant-$\alpha$ solution. We can remove the effect of cross-shore position from these expressions by subtracting $\tfrac{8}{9}\varepsilon x$ from $\overline{v}$ to get a function that is linearly dependent on $\varepsilon U_s^2$. Figure \ref{fig:Time-mean_x-compensated_alongshore_velocity} shows these new expressions and compares them with the data. We observe that the small-$\theta$, constant-$\alpha$ solution is more effective in capturing the average alongshore velocity compared with \citeauthor{ryrie1983longshore}'s analytic solution. This suggests that the analytic solution overestimates the alongshore velocity, especially at small $x$. The success of the small-$\theta$, constant-$\alpha$ solution is notable for collapsing breakers, given that the worst agreement between the theory and data occurred for these waves at station $S1$. For plunging breakers at larger $\varepsilon$ values, the alongshore velocity starts to exceed the $\Tilde{v}_C$ prediction. This could be due to the fact that the mean alongshore velocity is larger than its minimum value. Nevertheless, the data confirms that $\Tilde{v}_C$ can provide a reasonable prediction of the time averaged alongshore velocity during a swash cycle at different cross-shore positions for different wave breakers.

\section{Conclusions}
\label{sec:Conclusions}

We have extended the constant-$\alpha$ solution for one-dimensional swash flow due to a normally incident bore \citep{antuono2010shock} to find the alongshore swash flow solution using the weakly two-dimensional nonlinear shallow water equations introduced by \citet{ryrie1983longshore}. The weakly two-dimensional equations are derived under the assumption of a small effective bore approach angle, $\varepsilon$, leading to a one-way coupling from the cross-shore flow to the alongshore flow. Since we use the constant-$\alpha$ cross-shore flow to find the alongshore flow for small-$\theta$, we refer to our new solution as the `small-$\theta$, constant-$\alpha$' solution. A key distinction between our solution and previous analytic solution due to \citet{ryrie1983longshore} is that the characteristic variable associated with the alongshore flow, $\gamma$, is not constant throughout the swash zone. Rather, it rapidly decreases behind the moving shoreline.

The two parameters required to generate the solution are the constant-$\alpha$ value, which quantifies the strength of the bore, and the $\varepsilon$ value, which quantifies the bore's effective approach angle. While these parameters are based on the bore properties at the offshore boundary of the domain for the nonlinear shallow water equations, we show how measurements of the flow and shoreline motion in the swash zone can also be used to extract these parameters.   

We apply our new solution to data from large-scale experiments at various different wave conditions, including normally incident solitary waves and normally incident and obliquely incident regular waves with long periods to generate independent swash events. To generate obliquely incident waves, we use the wall-reflection method first introduced by \citet{dalrymple1989directional} as implemented in the AwaSys wave generation software \citep{meinert2011awasys}. This method is effective at minimizing the alongshore variability in the incoming waves, which would otherwise drive its own alongshore flow. Comparisons of our solution with experimental data confirm that the fundamental assumptions of theory are satisfied and that the theory predicts the flow field with good accuracy. The agreement between theory and data is good in the inner surf and swash zones, and even near the bore collapse location where previous analytic solutions were known not to be effective. From the theory, we also derive an explicit expression to the minimum alongshore velocity as a function of cross-shore distance, which shows good agreement with the data of the swash-cycle-averaged alongshore velocity, and could be used in predictive models of alongshore transport at coastlines.

As directions for future work, we recommend the following. We noted some sensitivity to how well the data match the theory for different breaker types, so future laboratory experiments could expand the present study to better understand this. Related to this, we have focused our analysis on the swash of a single bore free from interactions with other waves whereas future work could consider more realistic scenarios including these wave-swash interactions. We have also neglected the effects of bottom friction and energy losses due to turbulence, as well as the properties of turbulence in this flow. From our limited dataset, we observe a decrease in the mean flow energy with increasing onshore distance, suggesting that energy losses could be important. Finally, it is well known that Lagrangian mass transport velocities can exceed Eulerian mean flow velocities due to Stokes-drift-like effects in non-uniform oscillatory flows, and this deserves attention in future work to better quantify alongshore mass transport.

%\backsection[Supplementary data]{\label{SupMat}Supplementary material and movies are available at \\https://doi.org/10.1017/jfm.2019...}

\backsection[Acknowledgements]{H.S. appreciates the help of Dr. T. Maddux and R. Miller with the experimental setup and data acquisition at the Hinsdale Wave Research Laboratory at Oregon State University (Corvallis, OR, USA).}

\backsection[Funding]{The authors gratefully acknowledge the support from the Natural Hazards Engineering Research Infrastructure and National Science Foundation (OCE-2219845; OCE-2219846; CMMI-2037914).}

\backsection[Declaration of interests]{The authors report no conflict of interest.}

\backsection[Data availability statement]{The data that support the findings of this study are openly available in NSF NHERI DesignSafe, project ID: PRJ-4280.}

\backsection[Author ORCIDs]{H. Sung, https://orcid.org/0000-0002-4568-2370; P. Lomonaco, https://orcid.org/0000-0001-6721-5688; P. Chardon-Maldonado, https://orcid.org/0000-0001-9944-1495; R. P. Mulligan, https://orcid.org/0000-0002-2600-8647; J. Olsthoorn, https://orcid.org/0000-0002-3730-6156; J. A. Puleo, https://orcid.org/0000-0002-2889-5956; N. Pujara, https://orcid.org/0000-0002-0274-4527}

\backsection[Author contributions]{H.S.: Conceptualization; Conducting experiments; Data analysis and interpretation; Methodology; Theoretical analysis; Writing - original draft. P.L.: Conducting experiments; Methodology; Writing - review and editing. P.C.-M., R.P.M., J.O., J.A.P.: Conceptualization; Data analysis and interpretation; Funding acquisition; Project administration; Writing - review and editing. N.P.: Conceptualization; Data analysis and interpretation; Funding acquisition; Methodology; Theoretical analysis; Project administration; Supervision; Writing - review and editing.}

\appendix

\section{Evaluation of the numerical errors in alongshore flow solution}\label{sec:appA}

\begin{figure}
      \centerline{\includegraphics[width=1\linewidth]{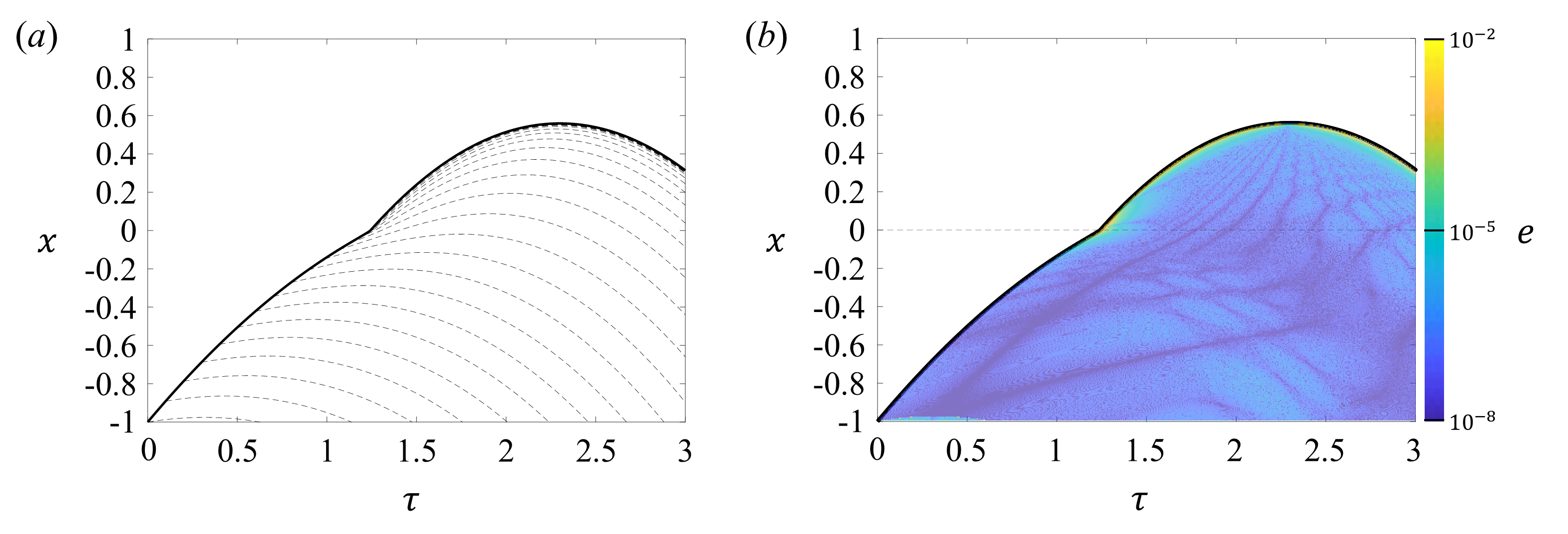}}
      \caption{(\textit{a}) $x_b$ (solid line) and the $\gamma_2$ characteristics (dashed lines) with $\alpha_2=2.3$ and $\varepsilon = 0.24$. (\textit{b}) The absolute error ($e$) of the NSWEs under shock solution.}
    \label{fig:xb_gamma2_absolute_error}
    \end{figure}

In this section, we evaluate how well the small-$\theta$, constant-$\alpha$ solution from \S \ref{sec:Numerically computed results} satisfies the governing equations. Figure \ref{fig:xb_gamma2_absolute_error}a shows the bore path $x_b$ and the $\gamma$ characteristics. We only show representative $\gamma$ characteristics; there are many more that were calculated, particularly near bore collapse due to the adaptive time step. The shoreline position after the bore collapse follows Eq.~\eqref{eq:Shen Meyer_x_s} and we can observe the transition to this ballistic solution at $x\approx0$. 
       
We evaluated the error, $e$, in the alongshore flow component by using the computed solution in the second-order accurate finite-difference scheme of the NSWEs (Eq.~\eqref{eq:NSWEs2c}) given by
\begin{equation}
    e = \left|\frac{v(i,j+1)-v(i,j-1)}{2\Delta\tau} +u(i,j)\frac{v(i+1,j)-v(i-1,j)}{2\Delta x} - \varepsilon\frac{h(i,j+1)-h(i,j-1)}{2\Delta\tau}\right| .
    \label{eq:error numerical scheme}
\end{equation}
Figure \ref{fig:xb_gamma2_absolute_error}b shows this error. While it is relatively high immediately behind the shoreline ($e = O(10^{-2}-10^{-1})$) and below the characteristics departing from the bore at $(x_b,\tau)=(-1,0)$ ($e = O(10^{-3})$) due to the imposed initial condition, these regions are small. For most of the flow field, the error is $O(10^{-8}-10^{-5})$. 

\bibliographystyle{jfm}
\bibliography{jfm}

\end{document}